\documentclass[12pt,preprint]{aastex}
\usepackage{url}
\newcommand{\mybf}{}
\newcommand{\mybftwo}{}
\shorttitle{HH objects and mid-infrared outflows in Vela C}
\shortauthors{Zhang et al.}

\begin{document}
\title{Herbig-Haro objects and mid-infrared outflows in the Vela C molecular cloud}
\author{Miaomiao Zhang\altaffilmark{1,2}, Hongchi Wang\altaffilmark{1}, and Thomas Henning\altaffilmark{3}}
\affil{$^1$Purple Mountain Observatory, \& Key Laboratory for Radio Astronomy, Chinese Academy of Sciences, 210008 Nanjing, PR China}
\affil{$^2$Graduate School of the Chinese Academy of Sciences, 100080 Beijing, PR China}
\affil{$^3$Max-Planck-Institut f\"{u}r Astronomie, K\"{o}nigstuhl 17, 69117 Heidelberg, Germany}
\email{miaomiao@pmo.ac.cn}

\begin{abstract}
We have performed a deep [SII] $\lambda\lambda$6717/6731 wide field Herbig-Haro (HH) object survey toward the Vela C molecular cloud with a sky coverage of about 2 deg$^2$. In total, 18 new HH objects, HH 1090-1107, are discovered and the two previously known HH objects, HH 73-74, are also detected in our [SII] images. {\mybf We also present an investigation of mid-infrared outflows in the Vela C molecular cloud using the \textit{Wide-field Infrared Survey Explorer} images taken from AllWISE data release. Using the method suggested by Zhang \& Wang, eleven extended green objects (EGOs) are identified to be the mid-infrared outflows, including 6 new mid-infrared outflows that have not been detected previously at other wavelengths and 5 mid-infrared counterparts of the HH objects detected in this work.} Using the {\mybf AllWISE} Source Catalog and the source classification scheme suggested by Koenig et al., we have identified {\mybf 56} young stellar object (YSO) candidates in the Vela C molecular cloud. The possible driving sources of the HH objects {\mybf and EGOs} are discussed based on the morphology of HH objects {\mybf and EGOs} and the locations of HH objects, {\mybf EGOs} and YSO candidates. {\mybf Finally we associate 12 HH objects and five EGOs with 10 YSOs and YSO candidates. The median length of the outflows in Vela C is 0.35\,pc and the outflows seem to be oriented randomly.}
\end{abstract}


\section{Introduction}
Mass outflows are ubiquitous and play an essential role in the process of star formation \citep{arc07,bal07}. The young stellar objects (YSOs) eject material while they are gathering mass through an accretion disk from the surroundings, which is proved to be an important way to transfer the excessive angular momentum from star-forming cores to the ambient interstellar medium \citep{shu94,sha07}. Outflows have been observed at different wavelengths {\mybf such as Herbig-Haro (HH) objects in the visual, molecular hydrogen emission line objects (MHOs) at 2.12\,\micron, and CO molecular outflows in the millimeter wavelength.}

HH objects are produced from the interaction of faster gas clumps with the slower ones that are ejected previously or the interaction of the supersonic gas ejected from YSOs with the ambient interstellar medium \citep{rei01}. During the cooling process of the shocked gas, HH objects can exhibit strong forbidden lines such as [SII] $\lambda\lambda$ 6717/6731 \citep{hollenbach89,hartigan99}. These lines are usually stronger than those in photo-ionized regions such as HII regions or planetary nebulae \citep{kwitter98,kwitter01}. Therefore, it is effective and efficient to survey the star forming regions using the [SII] $\lambda\lambda$ 6717/6731 narrowband together with a neighboring broadband for identifying HH objects.

The typical velocity of HH objects is about 100-300 km s$^{-1}$ and their typical dynamical age is about 10$^3$-10$^5$\,yr \citep{bal07}. Therefore, HH objects are good tracers of recent star formation. Mass outflows from YSOs are believed to be associated with the accretion process of protostars and to be driven magnetocentrifugally from open field lines anchored on rapidly rotating circumstellar disks \citep{kon00,shu00,sha07}. The identification of HH objects can provide hints for the locations of protostars, especially for the ones that are deeply embedded in the cloud cores. For example, a highly collimated jet usually points to its driving source and a pair of counterflows usually indicates that a YSO should lie between them. A wide-field survey of HH objects toward an entire star forming region can reveal the star formation activity in the star forming region. Taking advantage of the large field of view provided by mosaic CCDs, we have conducted a series of surveys with the aim to investigate the star formation activity in nearby star forming regions. Our surveys {\mybf have} covered the R CrA, Cha I, Lupus I, Lupus III and Vela C star forming regions. In this paper, we report the results from our survey toward the Vela C molecular cloud. Our results from other surveys can be found in \citet{wang04} and \citet{wh06,wh09}.

{\mybf Shocked gas in outflows has also abundant atomic and molecular hydrogen line emission in the mid-infrared. \citet{rosen00} obtained the infrared spectra from 2.4\,\micron~to 45\,\micron~toward the brightest H$_2$ emission peak of the Orion OMC-1 outflow using the Short-Wavelength-Spectrometer on the Infrared Space Observatory (ISO) and detected a wealth of emission and absorption features dominated by $\sim$50 H$_2$ ro-vibrational and pure rotational lines reaching from H$_2$ 0-0 S(1) to 0-0 S(25). Recent \textit{Spitzer} surveys toward star-forming regions using the InfraRed Array Camera \citep[IRAC,][]{irac} present an opportunity for the investigation of mid-infrared outflows. The IRAC 4.5\,\micron~band contains H$_2$ ($\nu$ =0-0, S(9, 10, 11)) lines and CO ($\nu$ =1-0) band heads \citep{reach06} and has proved an effective tool for exhibiting the mid-infrared emission from outflows \citep{nc04,harvey06,gal07,tei08,qiu08}. Some authors began to use the extended excessive emission at 4.5\,\micron~to identify the mid-infrared outflows \citep{ego08,cham09}. We also identified about 40 extended green objects (EGOs) as the mid-infrared outflows in Ophiuchus based on the extended excessive emission at 4.5\,\micron~using the \textit{Spitzer} archive data from \textit{c2d} legacy program \citep{zw09} and we proposed the excessive emission at 4.5\,\micron~as a powerful tool to search and investigate the mid-infrared outflows.}

{\mybf The Wide-field Infrared Survey Explorer \citep[WISE;][]{wise} launched on December 14, 2009 began its survey observations on January 7, 2010. WISE surveyed the entire sky in the 3.4, 4.6, 12, and 22\,\micron~mid-infrared bandpasses (W1, W2, W3, and W4) with a 40 cm telescope equipped with a camera containing four mid-infrared detectors that simultaneously imaged the same 47\arcmin$\times$47\arcmin~field-of-view on the sky. The spatial resolutions are 6.1\arcsec, 6.4\arcsec, 6.5\arcsec, and 12\arcsec~in the four bands (W1-W4), respectively. The sensitivity of WISE varies with position on the sky and improves toward ecliptic poles due to systematic changes in the depth of coverage and the zodiacal foreground emission. The estimated AllWISE sensitivities for low coverage sky away from the Galactic plane are about 0.054, 0.071, 0.73, and 5 mJy for the four bands, respectively. WISE uses similar passbands to \textit{Spitzer}: W1 and W2 have close correspondence to IRAC 1 and IRAC 2 while W4 compares well with MIPS 24\,\micron. 
Considering the similarities of passbands between WISE and \textit{Spitzer}, we apply the method that was suggested by \citet{zw09} to the WISE data and use the WISE data to investigate the mid-infrared outflows in the Vela C molecular cloud, which could offer a complementary sample of outflows in the highly obscured regions of Vela C.}

The Vela Molecular Ridge (VMR) is a large molecular cloud complex located along the Galactic plane, roughly between {\it l} $\approx$ 255\arcdeg \ and 275\arcdeg, {\it b} $\approx \pm$5\arcdeg, which consists of four separate clouds, A, B, C, and D \citep{mm91,pettersson08}. {\mybf \citet{lis92} carried out a careful distance determination from infrared photometry, star counting and comparison with other distance estimators and arrived at a distance 2 kpc for cloud B and 0.7 $\pm$ 0.2 kpc for cloud ACD, which confirmed that cloud B probably is a background cloud \citep{pettersson08}.} {\mybf The Balloon-borne Large Aperture Submillimeter Telescope (BLAST) is a balloon-borne 2 m telescope which has three wide bands at 250, 350, and 500\,\micron~\citep{pascale08}. The BLAST survey toward Vela covers about 50 deg$^2$ in the Vela Molecular Ridge\footnote{The BLAST map of Vela can be found at \url{http://blastexperiment.info/results.php}}. \citet{netterfield09} presented the results from this unbiased submillimeter Galactic survey and more than 1000 compact sources have been identified. \citet{netterfield09} estimated the physical parameters such as temperature, mass, and luminosity for each of these sources through fitting the three BLAST fluxes with a single-temperature modified blackbody SED. They found that the temperatures of these BLAST sources are in a range of only several Kelvin up to 30K. }

{\mybf Much work has been done toward Vela D, including the identification of clusters and isolated protostars \citep{massi00,massi03,gian07,stra10} and the identification of jets and outflows \citep{wb99,elia07,loren02,gian05,gian13}. \citet{yam99b} suggested that both Vela C and D have recent star forming activity and the Vela C is less evolved than Vela D. The Vela C molecular cloud is attracting an increasing interest since it was selected as target of recent surveys such as Herschel\footnote{\url{http://www.herschel.fr/cea/hobys/en/index.php}} and ESO-LABOCA \citep{massiposter}. The mass of Vela C estimated from CO} is about 3$\times$10$^5$ M$_{\odot}$ \citep{mm91}. \citet{yam99b} also detected 13 dense C$^{18}$O clumps in the Vela C molecular cloud. {\mybf \citet{hill11} and \citet{gian12} presented the Herschel PACS and SPIRE survey of the Vela C. \citet{hill11} obtained the temperature map of the Vela C and found that the dust temperatures in most of the Vela C region are significantly lower than 30K. \citet{gian12} extracted the point sources on the Herschel maps and they found that the mean temperature of protostellar sources is 12.8 K which is warmer than that for the starless sources of 10.3 K. These all suggest that the Vela C molecular cloud may stay at a early evolutionary stage of star formation.}

More than 80 YSOs have been identified in the Vela C molecular cloud through H$\alpha$ imaging, near-infrared imaging and spectroscopy observations \citep{pre94,massi03,bt04,bik05,tb05,baba06,pettersson08}. The outflows detected in Vela C include a confirmed CO molecular outflow \citep{yam99b} and six HH objects, i.e., HH 73-75, HH 133, and HH 1042-1043. Among them, HH 75 and HH 133 are located in the northeast of Vela C, which is outside the region harboring the dense C$^{18}$O clumps. HH 1042 and HH 1043 are two jets located in the massive star-forming region RCW 36 that are identified using X-shooter on the ESO {\it Very Large Telescope} \citep{eller11,eller13}.  More details about the Vela C molecular cloud can be found in the review of \citet{pettersson08}.

{\mybf \section{Data and Analysis}}
{\mybf \subsection{WFI data}}
The observations were conducted on 2007 February 24-28 and March 01-02 using the European Southern Observatory Max Planck Gesellschaft (ESO MPG) 2.2 m telescope at La Silla. The telescope is equipped with the Wide Field Imager (WFI) camera which is a mosaic of eight 2k$\times$4k CCDs with narrow interchip gaps (the filling factor is 95.9\%; \citealt{baade99}). With a plate scale of 0\farcs238 pixel$^{-1}$, the field of view of WFI is 34\arcmin$\times$33\arcmin. The combination of the filters ESO 857 and ESO 847 was used for the observations. The ESO 857 filter is a narrow-band filter with a central wavelength of $\lambda_{c} =$ 6763.4\,\AA \ and a bandpass of $\Delta\lambda =$ 84.2\,\AA. Its transmission at the characteristic lines of HH objects, [SII] $\lambda\lambda 6717/6731$, is 0.31 and 0.63, respectively. The ESO 847 filter that is used to measure the continuum emission  is an intermediate-band filter with a central wavelength of $\lambda_{c} =$ 7217.9\,\AA \ and a bandpass of $\Delta\lambda =$ 256.5\,\AA. The obtained data were reduced using the MSCRED package in IRAF\footnote{IRAF is distributed by the National Optical Astronomy Observatory, which is operated by the Association of Universities for Research in Astronomy, Inc., under cooperative agreement with the National Science Foundation.} in the same way as in \citet{wang04} and \citet{wh06,wh09}.

\begin{figure}
\includegraphics[width=15cm]{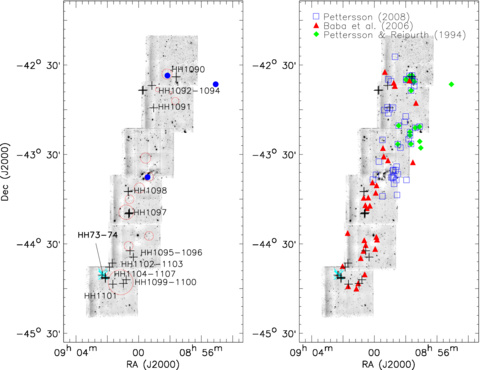}
\caption{Mosaic [SII] images of the Vela C molecular cloud. The observed fields 1-6 are from lower left to upper right. {\mybf The  \textit{left} panel} shows the locations of HH objects in this paper. The newly detected HH objects, HH 1090-1107, are marked with pluses and the two previously known HH objects, HH 73 and 74 \citep{reipurth88}, are labeled with {\mybf cyan crosses}. The C$^{18}$O dense clumps identified by \citet{yam99b} are marked with red circles whose sizes are in proportion to the physical sizes of the cores. {\mybf The blue filled circles show the locations of known young embedded star clusters identified by \citet{massi03}}; {\mybf the \textit{right} panel} shows the locations of known YSOs {\mybf and YSO candidates} in the Vela C molecular cloud. The {\mybf green filled diamonds} mark the locations of YSOs from \citet{pre94}. The {\mybf red filled triangles} mark the locations of protostars from \citet{baba06}, while the {\mybf blue empty squares} mark the locations of pre-main sequence objects and candidates from \citet{pettersson08}.\label{fig:mosaic1}}
\end{figure}

In total, six fields toward the Vela C molecular cloud were observed (see Fig.~\ref{fig:mosaic1}). To eliminate the effects of interchip gaps, cosmic rays, and bad pixels, six frames were typically taken in each of the ESO 857 and ESO 847 filters with the telescope pointing being slightly offset. Note that fields 2-4 were observed with only four frames in the ESO 857 filter. The exposures in the ESO 857 and ESO 847 filters for each frame were 1400 and 450s, respectively. The astrometry for our resultant images was done using the USNO2 catalog. It was found that our astrometry is accurate within 0\farcs35 both in R.A. and declination.

{\mybf \subsection{Identification of HH objects}}
{\mybf For each field, we have obtained the [SII] narrow-band image and the continuum intermediate-band image. The \textit{daophot} package of IRAF is used to detect point sources on [SII] images and continuum images. Then the unsaturated bright point sources are used to adjust the flux level of stars in the [SII] image and the continuum image and finally we obtain a continuum-subtracted [SII] image for each field.}

{\mybf All continuum-subtracted [SII] images have been visually inspected for extended excessive emission features. These extended excessive emission features are treated as HH candidates. Every HH candidate has been examined in the corresponding [SII] and continuum images in order to avoid the inclusion of instrumental artifacts. Any object that could be an potential artifact has been removed from the HH candidate list. Emission nebulae, such as planetary nebulae (PNe), supernova remnants (SNRs) and HII regions may contaminate our identification of excessive [SII] emission objects to be HH objects. PNe usually exhibits symmetrical morphology and can be distinguished from HH objects based on their morphological difference. {\mybftwo We found no PNe in our HH candidate list.} We also searched the known SNRs and HII regions in the \textit{SIMBAD} database and examined HH candidates near these objects carefully. {\mybftwo We found that three ($\sim$13\%) HH candidates are associated with the HII region, RCW 36, and we removed them from our HH candidate list.} 
The remaining extended emission features are identified as HH objects and listed in table~\ref{tab1}. The official numbers of HH objects are supplied by Bo Reipurth who is in charge of the assignment of official HH object numbers on behalf of IAU\footnote{\url{http://www.iau.org/}}.}

{\bf \subsection{Photometry of the features of HH objects}}
{\mybf Aperture photometry is applied to the continuum-subtracted [SII] images for each HH object feature. We define an elliptic aperture around each HH feature in the continuum-subtracted [SII] images based on the morphology and surface brightness distribution of the HH feature. The principle for the aperture definition is to ensure that no stars are included in the aperture and the aperture contains as little area that seems to be of no excessive [SII] emission as possible. In the same time, we also define a circular nearby sky aperture in the emission-free region to estimate the local sky level. The elliptic apertures that we define for each HH feature are shown in Figs.~\ref{fig:nhh1}-\ref{fig:hh} with white ellipses. Then we measure the flux for each HH feature in units of {\mybftwo digital number (DN)}.}

{\mybf Flux calibration is conducted using the UCAC4 catalog \citep{ucac4}, which contains precise and reliable photometry in $r^{\prime}$ (Sloan filter) band taken from the sixth data release (DR6) of the AAVSO Photometric All-Sky Survey (APASS\footnote{\url{http://www.aavso.org/apass}}). Sextractor \citep{sex} is used to perform photometry on the [SII] images. The conversion from DN to Jy is done through comparing the instrumental fluxes of stars in the bandpass of the [SII] filter with the cataloged $r^{\prime}$-band fluxes in the UCAC4 catalog. Note that here we assume for each star that the specific intensity in the Sloan $r^{\prime}$ band is equal to that in the WFI [SII] band. The transmittance of the WFI [SII] filter at the wavelengths of $\lambda$6717 and $\lambda$6731 is 0.31 and 0.63, respectively, which are different by a factor of $\sim$2. Therefore, an accurate measurement of the total fluxes in the [SII] duplicate lines requires knowledge of the strength ratio between the two emission lines, which however can not be determined from our imaging data. Spectroscopical observations has shown that the [SII] $\lambda\lambda$6717 to $\lambda\lambda$6731 line ratios for HH objects lie typically in the range of 0.6 to 1.4 \citep{aya00,riera01,wang01,wang03}. For the purpose to estimate the total fluxes in the [SII] $\lambda\lambda$ 6717/6731 lines of HH objects and for simplicity, we assume that the [SII] $\lambda$ 6717 flux is equal to the [SII] $\lambda$ 6731 flux. With this assumption and taking into account the transmission profile of the WFI [SII] filter, we can transform the fluxes of HH features from the units of DN to the flux units of W m$^{-2}$. The results of photometry for HH features are listed in table~\ref{tab1}. In table~\ref{tab1} we also provide the peak flux in the elliptic aperture and its signal to noise ratio for each HH feature.}

{\mybf Uncertainties in the photometry are mainly caused by the variation of the local background level in the continuum-subtracted [SII] images and choice of the sizes of the elliptic apertures. We estimate that our detection limit of [SII] $\lambda \lambda$6717/6731 emission is 2.7$\times$10$^{-20}$ W\,m$^{-2}$\,arcsec$^{-2}$ (3$\sigma$), which is similar to our [SII] surveys toward other star-forming regions, i.e., 2.3$\times$10$^{-20}$ W\,m$^{-2}$\,arcsec$^{-2}$ (3$\sigma$) for the survey toward Cham I cloud \citep{wh06} and 3.3$\times$10$^{-20}$ W\,m$^{-2}$\,arcsec$^{-2}$ (3$\sigma$) for the survey toward the Lupus I and III clouds \citep{wh09}.}

{\mybf \subsection{WISE data}}
{\mybf The WISE data used in this work are taken from the AllWISE data release\footnote{\url{http://wise2.ipac.caltech.edu/docs/release/allwise/}}, which combines data from the cryogenic and post-cryogenic survey phases to form the most comprehensive view of the mid-infrared sky currently available. Released data products include an atlas of coadded image sets and the source catalogs. The details about the WISE data acquisition and reduction can be found in \citet{wise}, \citet{jarrett11} and Explanatory Supplement to the AllWISE Data Release Products\footnote{\url{http://wise2.ipac.caltech.edu/docs/release/allwise/expsup/}}.}

{\mybf We downloaded the WISE data from NASA/IPAC Infrared Science Archive \footnote{\url{http://irsa.ipac.caltech.edu}} and restricted the query region to RA between 08:54:00 to 09:04:00 and Dec between -45:30:00 to -42:00:00. The mosaic three-color images constructed with W1 (blue), W2 (green), and W3 (red) images is shown in Fig.\ref{fig:mosaic2}. The raw images were cut into many subregions of 10\arcmin$\times$10\arcmin. In each subregion, the convolution kernel built by \citet{kernels} is used to convolve the W1 band images so that the images at 3.4\,\micron~and 4.6\,\micron~have the same PSF. Then we use the IDL routines \textit{find} and \textit{aper} to do the photometry on the W2 band images and the convolved W1 band images. After matching our detected sources with the AllWISE source catalog, the quality flags in the AllWISE source catalog are used to select unsaturated point sources in our detected source sample. We use these stars to estimate the average flux ratio of stars between the W1 and W2 bands. This flux ratio is used to scale the flux level of convolved W1 band image. Then we obtained the resultant difference images between W1 and W2 bands, which are the subtraction of emission in band 1 from that in band 2. We also obtained three-color images constructed using W1 (3.4\,\micron, blue), W2 (4.6\,\micron, green), and W3 (12\,\micron, red) bands for each subregion. These difference and three-color images are used to identify mid-infrared outflows in the Vela C molecular cloud.}

{\mybf \subsection{Identification of WISE mid-infrared outflows}}
{\mybf Although the WISE passbands are similar to the \textit{Spitzer} passbands, there are still some differences between them. For example, {\mybftwo W1 passband is slightly bluer than IRAC 1 band while W2 passband is slightly redder than IRAC 2 band.} Moreover, W3 is a unique passband which is different from IRAC 4 but comparable to the IRAS 12\,\micron~band. Compared with IRAC 4, W3 is redder and  sensitive to both PAH emission and amorphous silicate absorption (10\,\micron). Therefore, we need to examine whether the method suggested by \citet{zw09} can be applied to the WISE data. For this purpose, we have inspected the EGOs identified by \citet{zw09} in the corresponding WISE images.}

{\mybf \citet{zw09} identified 44 EGOs that consist of 122 EGO features in Ophiuchus using \textit{Spitzer} IRAC data. We checked each of EGOs in the WISE images and found that 95 EGO features contained in 34 EGOs show similar extended excessive emission in the difference images between W1 and W2 bands. They also exhibit green color in the three-color images constructed with W1, W2, and W3 bands. However, 10 faint EGOs are missing in WISE images due to the poor sensitivity of WISE. 
Figure~\ref{fig:ophego} shows the example of EGO 41 and 43 in Ophiuchus, which exhibit  similar structures in \textit{Spitzer} and WISE images. Thus the method suggested by \citet{zw09} is also effective with the WISE data and can be used to search for mid-infrared outflows.}

{\mybf For the Vela C molecular cloud, the WISE mosaics of the difference image between W1 and W2 band were displayed and visually inspected to search for mid-infrared outflows based on appearance of extended excessive emission at 4.6\,\micron~(W2). All extended excessive emission features are selected as the EGO candidates. This is the preliminary identification. Subsequent cleaning of these EGO candidates is performed in following two stages.}

{\mybf Firstly, all EGO candidates are examined in the corresponding W1-W4 band images and the AllWISE source catalog to exclude artifacts and point sources. In this stage, the EGO candidates with HH counterparts are all reserved. But for the EGO candidates without HH counterparts, only the extended objects are reserved. Then the surviving EGO candidates have been further confirmed as bona fide EGOs using the three-color images constructed with W1 (blue), W2 (green), and W3 (red) bands.}

{\mybftwo Secondly, reflection nebulae, PNe, SNRs, and HII regions may contaminate our EGO sample. PNe can be distinguished from EGOs based on their symmetrical morphology while reflection nebulae can be identified based on their extended emission in R band or K band. We have checked each EGO in the DSS2 red image and 2MASS Ks image to exclude reflection nebulae. We also searched the known SNRs and HII regions in \textit{SIMBAD} database and examined EGOs near them.} Any feature that is suspected to be associated with these contaminants has been removed from our list of EGOs.

{\mybf Finally we obtain 11 EGOs as the mid-infrared outflows in the Vela C molecular cloud. Their coordinates, number of features, and association with HH objects are {\mybftwo listed} in Table~\ref{egotable}.}

{\mybf \subsection{Photometry for the EGO features}}
{\mybf Areal photometry that is suggested by \citet{ego08} was performed on all EGO features. We define an arbitrarily shaped polygonal aperture for each EGO feature to closely match its morphology in 4.6\,\micron. Great care is taken to ensure that no stars are included in the apertures and the apertures contain as little as possible emission-free region. A nearby circular aperture is also defined to estimate the local sky level. The polygonal apertures and sky apertures are applied to flux density measurements in WISE W1-W3 bands. Note that we did not measure the flux densities in WISE W4 band for EGO features due to the large beam size at 22\,\micron. The polygonal apertures that we defined for photometry of EGO features are shown in Figs.~\ref{fig:nhh1}-\ref{fig:ego9} with white polygons.}

{\mybf Uncertainties in the photometry are calculated based on the variation of the local sky level in the corresponding WISE images and choice of sizes of the polygonal apertures. The results of photometry are reported in Table~\ref{egophoto}. }

{\mybf \section{Results and Discussion}}
{\mybf \subsection{HH objects in Vela C}}
With a total sky coverage of $\sim$2 deg$^2$, our deep [SII] survey embraces the main region of the Vela C molecular cloud, in which 13 C$^{18}$O (1-0) dense clumps have been detected by \citet{yam99b}. The six observed fields toward Vela C are shown from lower left to upper right in Fig.~\ref{fig:mosaic1}.

In total 20 HH objects are detected, among which 18 HH objects, i.e., HH 1090-1107, are newly discovered. We also detected two previously known HH objects, HH 73-74. {\mybf For the remaining known HH objects in the region, HH 75 and HH 133 are not located in the coverage of our survey. 
{\mybftwo HH 1042-1043 are spatially resolved in [Fe II] line map by VLT/SINFONI \citep{eller11}. The X-shooter spectroscopic observations of these two jets cover the wavelength range of 3000 \AA~to 2.5\,\micron~and reveal the optical emission lines such as H$\alpha$~$\lambda$6563\AA~and [SII]~$\lambda\lambda$6717/6731\AA~in the jets, which results in the inclusion of these two jets in the catalog of HH objects \citep{eller11,eller13}. We also checked them in our continuum-subtracted [SII] images and found that they are not detected due to the bright emission from the HII region, RCW 36.}} The coordinates and morphologies of the detected HH objects are listed in table~\ref{tab1}. Figure~\ref{fig:mosaic1} shows the locations of the HH objects detected in our survey. The newly discovered HH objects, HH 1090-1107, are marked with {\mybf black pluses} and the two previously known HH objects, HH 73 and 74, are labeled with {\mybf cyan crosses}. {\mybf 13 C$^{18}$O clumps identified by \citet{yam99b} are marked with red circles whose scales are in proportion to the sizes of C$^{18}$O clumps. We also show the locations of known young embedded star clusters from \citet{massi03} with blue filled circles in Fig.~\ref{fig:mosaic1}.} We can see that the detected HH objects are mainly concentrated in the north and south region of the Vela C molecular cloud. 


{\mybf \subsection{Mid-infrared outflows in Vela C}}
{\mybf We have identified six new WISE mid-infrared outflows and five mid-infrared counterparts of known HH objects in the Vela C molecular cloud. Their IDs, coordinates, and photometry are list in Table~\ref{egotable} and \ref{egophoto}. The EGOs in table~\ref{egotable} are numbered by order in right ascension. About the nomenclature of EGOs, we refer to the work of \citet{davis10}\footnote{\url{http://www.astro.ljmu.ac.uk/MHCat/definition.html}}. The essential principle is that wherever possible, the whole outflow should be given a single EGO name.  If it is not clear whether the widely separate objects are associated with the same outflow, separate EGO names could be given to each feature. Figure~\ref{fig:mosaic2} shows the spatial distribution of EGOs in Vela C with white circles. We can see that EGOs are mainly concentrated on the dense region of Vela C which harbours the C$^{18}$O clumps \citep{yam99b}.}

\subsection{Young stellar population in Vela C}

To search for the possible driving sources of HH objects {\mybf and EGOs}, we first compile a catalog of YSOs in the Vela C molecular cloud. 

{\mybf \subsubsection{Known YSOs and YSO candidates from literature}\label{sect:knownyso}}

\begin{figure}
\includegraphics[width=15cm]{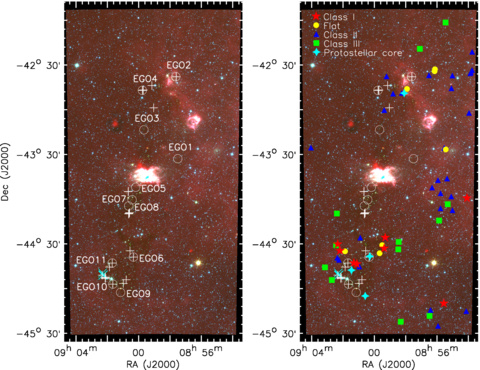}
\caption{{\mybf Mosaic three-color images of the Vela C molecular cloud which are constructed with WISE 3.4\,\micron~(blue), 4.6\,\micron~(green), and 12\,\micron~(red) images.} {\mybf The \textit{left} panel} shows the locations {\mybf of mid-infrared outflows identified in this paper with white circles. The newly discovered HH objects are marked with white pluses and the previously known HH objects, HH 73 and HH 74, are labeled with cyan crosses. The \textit{right} panel shows the locations of YSOs and YSO candidates identified using AllWISE source catalog with red stars (Class I), yellow filled circles (``Flat spectrum"), blue filled triangles (Class II), and green filled squares (Class III), respectively. The new protostellar core candidates identified using BLAST cores \citep{netterfield09} and AllWISE source catalog are also labeled with cyan stars.} \label{fig:mosaic2}}
\end{figure}

\begin{figure}
\includegraphics[width=15cm]{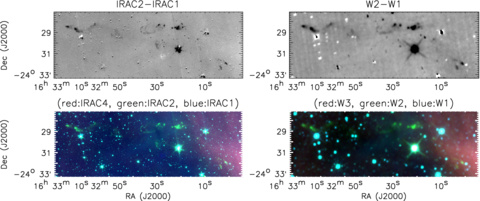}
\caption{{\mybf Region of EGO 41 and 43 that are identified by \citet{zw09} in Ophiuchus. The images are the difference image of IRAC channel 2 (4.5\,\micron) minus channel 1 (3.6\,\micron) (\textit{top-left}); the difference image of WISE channel 2 (4.6\,\micron) minus channel 1 (3.4\,\micron) (\textit{top-right}); the three-color image constructed with IRAC 3.6\,\micron~(blue), 4.5\,\micron~(green), and 8.0\,\micron~(red) bands (\textit{bottom-left}); and the three-color image constructed with WISE 3.4\,\micron~(blue), 4.6\,\micron~(green), and 12.0\,\micron~(red) bands (\textit{bottom-right}), respectively.}}
\label{fig:ophego}
\end{figure}

{\mybf \citet{pre94} {\mybftwo have} presented an H$\alpha$ emission line star sample based on their objective prism survey toward the Vela Molecular Ridge. Using the data of deep near-infrared survey toward Vela C, \citet{baba06} identified 31 protostar candidates in Vela C based on the coincidence between the near-infrared point sources and the mid-infrared MSX and/or IRAS sources. \citet{pettersson08} reviewed the star formation in Vela Molecular Ridge and compiled a list of spectroscopically confirmed pre-main sequence objects (PMS) and candidates. We matched these three catalogs with a tolerance of 1\arcsec~and finally obtained 83 YSOs and YSO candidatas in the region that is shown in Fig.~\ref{fig:mosaic1} and Fig.~\ref{fig:mosaic2}. Note especially that we do not include the protostellar cores identified by \citet{gian12} in our known YSO sample because the complete Herschel catalogue of Vela-C has not been released.}

Figure~\ref{fig:mosaic1} shows the locations of the known YSOs and YSO candidates in Vela C. The {\mybf green filled diamonds} mark the locations of the H$\alpha$ emission line stars from \citet{pre94}. The {\mybf red filled triangles} mark the locations of the protostars that are identified by \citet{baba06} while the {\mybf blue empty squares} mark the locations of the spectroscopically confirmed pre-main sequence objects (PMS) and candidates from \citet{pettersson08}.

{\mybf \subsubsection{New YSO candidates identified using AllWISE source catalog}\label{sect:wiseyso}}

The known YSOs in Vela C are mainly identified based on optical or near-infrared observations. In order to investigate the YSOs that are deeply embedded in clouds, we use {\mybf the AllWISE Source Catalog.} We define a region of R. A. from 08$^h$54$^m$00$^s$ to 09$^h$04$^m$00$^s$ and declination from -45\arcdeg30\arcmin00\arcsec~to -42\arcdeg00\arcmin00\arcsec~{\mybf (see Figs.~\ref{fig:mosaic1} and \ref{fig:mosaic2})} to search for {\mybf YSOs in the AllWISE} Source Catalog. 

{\mybf Our unbiased search in the AllWISE source catalog returns over 130, 000 sources. 
We apply some initial cuts to the raw sources using the quality flags in the AllWISE source catalog:} 
\begin{itemize}
\item[1.] {\mybf Using the raw source catalog as the input and excluding extended sources (ext\_flg\footnote{\url{http://vizier.u-strasbg.fr/viz-bin/VizieR-n?-source=METAnot&catid=2328&notid=4&-out=text}}$\neq$0), the output is a point source catalog, {\mybftwo including $\sim$110, 000 point sources.}}

\item[2.] {\mybf Using the point source catalog as the input and excluding saturated sources (saturated pixel fraction in any WISE band\footnote{\url{http://cdsarc.u-strasbg.fr/viz-bin/Cat?II/328}} is not zero), the output is an unsaturated point source catalog, {\mybftwo including $\sim$100, 000 unsaturated point sources.}}

\item[3.] {\mybf Using the unsaturated point source catalog as the input and excluding contamination or confusion artifacts (cc\_flags\footnote{\url{http://vizier.u-strasbg.fr/viz-bin/VizieR-n?-source=METAnot&catid=2328&notid=3&-out=text}}$\neq$`0000'), the output is a high quality point source catalog (HQcatalog, hereafter), {\mybftwo including $\sim$43, 000 high quality point sources.}}
\end{itemize}

{\mybf Our identification of YSOs is based on our HQcatalog. Note that HQcatalog is a biased sample.}

\begin{figure}
\epsscale{1.0}
\plotone{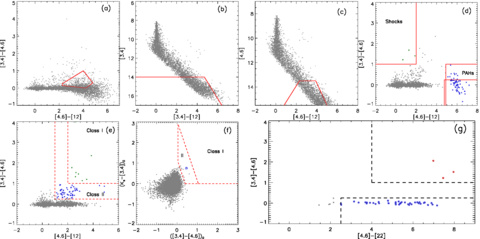}
\caption{Color-color diagrams that are used in the process of \textit{WISE} source classification in the Vela C molecular cloud. We use {\mybf AllWISE} Source Catalog to identify the YSO candidates in Vela C, based on the color criteria suggested by \citet{koenig12}. (a)-(e) show the phase 1 scheme which focuses on the sources that are detected at 3.4, 4.6, and 12\micron, requiring photometric uncertainty $<$0.2 mag in each of the three bands; the possible star-forming galaxies (a), the unresolved broad-line AGNs (b and c), and the shocks and resolved PAH emission objects (d) are selected and rejected as contaminations. Then the Class I and Class II sources are identified using the criteria shown in (e). (f) shows the phase 2 scheme which focuses on the sources that are detected at 3.4 and 4.6\micron, but not at 12\micron. We match these sources with the 2MASS All-Sky Point Source Catalog and deredden the {\it WISE}$+$2MASS photometry based on the extinction law presented in \citet{flaherty07}. Using the dereddened colors, YSO candidates are selected with the criteria shown in (f). {\mybf (g) shows the phase 3 scheme which focuses on the remaining sources from phase 1 and phase 2, that have detections at 22\,\micron~(SNR $>$ 10 and photometric uncertainties $<$ 0.2). Using the color criteria shown in (g), the transition disks (blue circles) and additional Class I sources (red circles) are selected.}\label{fig:ccd}}
\end{figure}

We {\mybf adopt} the source classification scheme in \citet{koenig12} {\mybf to identify YSO candidates in Vela C. This scheme consists of three ordinal stages and is established by applying the source classification scheme of \citet{gutermuth08,gutermuth09} to the {WISE}  wavebands.} Figure~\ref{fig:ccd} shows the color-color diagrams used in the process of YSO candidate identification. The details of this source classification scheme can be found in the appendix of \citet{koenig12}. Here we just summarize our process:

{\mybf Phase 1: for {WISE} sources with detections at 3.4, 4.6, and 12\,\micron, we restrict our sources to those with photometric uncertainty $<$ 0.2\,mag in each of the three bands. Then the possible star-forming galaxies (Fig.~\ref{fig:ccd}a), unresolved broad-line AGNs (Fig.~\ref{fig:ccd}b,c), shock objects and resolved PAH emission objects (Fig.~\ref{fig:ccd}d) are selected and rejected as the contaminants. After applying the color criteria suggested by \citet{koenig12} to the remaining sources, eight Class I sources and 58 Class II sources are selected (Fig.~\ref{fig:ccd}e).}

{\mybf Phase 2: for the {WISE} sources with detections at 3.4 and 4.6\,\micron~but without detections at 12\,\micron, we match them with 2MASS JHKs point source catalog \citep{2mass} using a tolerance of 1\arcsec~and remove the objects that have the bad photometric quality flags in the 2MASS catalog. Then we use the extinction map from \citet{dobashi11} to deredden the {WISE}$+$2MASS photometry based on the extinction law presented in \citet{flaherty07}. Using the dereddened colors, one Class II {\mybftwo source is} selected (see Fig.~\ref{fig:ccd}f).}

{\mybf Phase 3: this phase uses the remaining sources from phase 1 and phase 2 as the input, from which we select the sources that are detected at 22\,\micron~with SNR $>$ 10 and photometric uncertainty $<$ 0.2\,mag. Using the color criteria shown in Fig.~\ref{fig:ccd}g, 39 transition disks and three Class I sources are identified. The previously classified Class I sources in phases 1, 2, and 3 and Class II sources in phases 1 and 2 are also re-checked against highly reddened Class II sources and stars with disks and finally we remove one star with disk from our YSO candidate list.}

{\mybf After the above source classification, we have obtained 108 YSO candidates, including 11 Class I sources, 58 Class II sources, and 39 transition disks.} For these {\mybf 108} objects, we inspect the images in all four {WISE} bands and we find that there are some wrong detections due to the bright extended background emission in band 3 (12\,\micron) {\mybf and 4 (22\,\micron)}. After removing the objects with wrong photometry at 12\,\micron~{\mybf and 22\,\micron}, {\mybf 11} Class I sources, {\mybf 37} Class II sources, {\mybf and 8 transition disks} remain as YSO candidates.

{\mybf We match these 56 YSO candidates with the known YSOs obtained in section~\ref{sect:knownyso} with the tolerance of 1\arcsec~and we find that all 56 YSO candidates are newly discovered. Table~\ref{ysotable} lists these 56 new YSO candidates.}

\begin{figure}
\epsscale{0.7}
\plotone{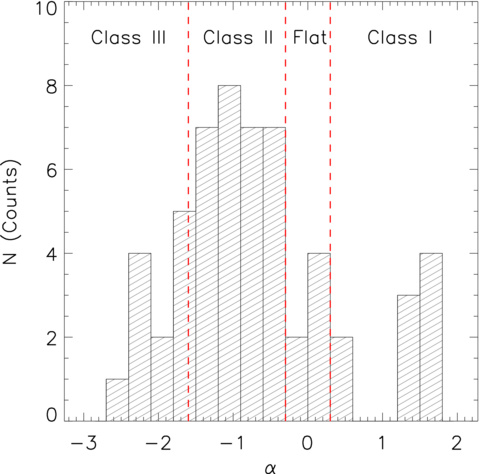}
\caption{{\mybf Histogram of the observed spectral indices of the 56 YSO candidates identified using AllWISE source catalog. The red dashed lines show the criteria of YSO classification scheme suggested by \citet{greene94}.}}
\label{fig:alpha}
\end{figure}

{\mybf The spectral index, defined as}
\begin{displaymath}
{\mybf \alpha = \frac{dlog(\lambda S(\lambda))}{dlog(\lambda)}}
\end{displaymath}
{\mybf where $S(\lambda)$ is the source's flux density at wavelength $\lambda$, can be used to indicate the evolutionary state of a source. We calculate the spectral indices of all 56 YSO candidates through fitting their observed SEDs from 2\,\micron~(2MASS Ks band) to 22\,\micron~(WISE W4 band). Using the YSO classification scheme suggested by \citet{greene94}, we reclassify these 56 YSO candidates into 9 Class I sources, 6 ``Flat spectrum" sources, 29 Class II sources, and 12 Class III sources. Figure~\ref{fig:alpha} shows the histogram of spectral indices of YSO candidates and Fig.~\ref{fig:mosaic2} shows the spatial distribution of YSO candidates in the Vela C molecular cloud.}

{\mybf \subsubsection{Protostellar core candidates in Vela C}} 

{\mybf A BLAST core with mid-IR or far-IR counterparts is likely to be a protostellar core. Thus we use the association between BLAST sources and WISE sources to identify protostellar core candidates in the Vela C molecular cloud. Note that here we just use the WISE sources with detections at 22\,\micron~to search for BLAST cores that harbor deeply embedded sources}.

{\mybf The diffraction-limited optics are designed to provide BLAST with a resolution of 30\arcsec, 42\arcsec, and 60\arcsec~at the three wave bands, respectively \citep{pascale08}. The radiation is imaged onto the detector arrays using a pair of cooled off-axis parabolic mirrors arranged in a ``Gaussian beam telescope" configuration \citep{devlin01}. The FWHMs of beams in the 250, 350, and 500\,\micron~bands are 36\arcsec, 42\arcsec, and 60\arcsec, respectively \citep[BLAST06, ][]{netterfield09}. The in-flight pointing of BLAST is accurate to $\sim$30\arcsec, and postflight pointing reconstruction is accurate to better than 5\arcsec~\citep{pascale08}. \citet{netterfield09} estimated the sizes of BLAST cores with Gaussian fit and found that the typical size of the BLAST cores in VMR is 62\arcsec. We finally adopt a tolerance of 36\arcsec~to match BLAST cores with WISE sources, which can produce ``reliable" associations but could miss some WISE sources that are associated with the large BLAST cores.}

{\mybf In the region shown in Figs.~\ref{fig:mosaic1} and \ref{fig:mosaic2}, there are 249 BLAST sources.} We match these {\mybf 249} BLAST sources with {\mybf our HQcatalog} using a tolerance of 36\arcsec. {\mybf We further restrict our sample to the sources that are detected at 22\,\micron~with SNR$>$3.} As a result, we {\mybf obtain 14} BLAST cores that are associated with {WISE} 22\,\micron~sources in the Vela C molecular cloud. {\mybf We then match these 14 BLAST cores with our known YSO catalog (section~\ref{sect:knownyso}) and the YSO catalog identified using AllWISE source catalog (section~\ref{sect:wiseyso}). We exclude the cores that are associated with known YSOs and YSOs identified using AllWISE source catalog to search for BLAST cores that are in the very early stage of star forming. Thus we obtain 10 BLAST protostellar core candidates.} 

For these 10 objects, we inspect the images in all four {WISE} bands and reject the sources with wrong detections at 22\,\micron. Finally {\mybf four} BLAST sources associated with {WISE} 22\,\micron~sources are obtained and we propose these objects to be the new protostellar core candidates in the Vela C molecular cloud. Table~\ref{blastproto} lists these {\mybf four} protostellar core candidates. The {\mybf cyan stars} in Fig.~\ref{fig:mosaic2} mark the locations of these {\mybf four} protostellar core candidates.


{\mybf \subsection{Possible driving sources of HH objects and EGOs in Vela C}}
{\mybf Based on the morphologies of HH objects and EGOs and the locations of HH objects, EGOs, and YSOs, we investigate the possible driving sources of HH objects and EGOs. Finally we associate 12 HH objects and five EGOs with 10 YSOs and YSO candidates. Table~\ref{extable} lists these 10 YSOs and their associated HH objects and EGOs. Due to the lack of stronger evidence such as proper motions, the identification of driving sources has certain uncertainty. Actually, only the driving source of HH 1098 can be confirmed to be IRS 35 because HH 1098 is associated with a CO outflow driven by IRS 35 \citep{yam99b}. Details about the possible driving sources of HH objects and EGOs can be found in section~\ref{sect:notes}.}

\begin{figure}
\epsscale{0.8}
\plotone{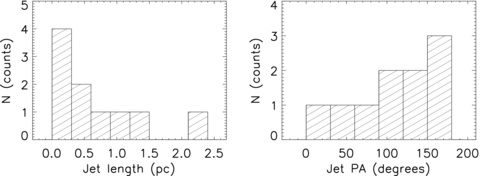}
\caption{{\mybf Histograms of jet lengths in bins of 0.3\,pc (\textit{left}) and jet position angles in bins of 30\degr~(\textit{right}).}}
\label{fig:dist}
\end{figure}

\begin{figure}
\epsscale{1.0}
\plotone{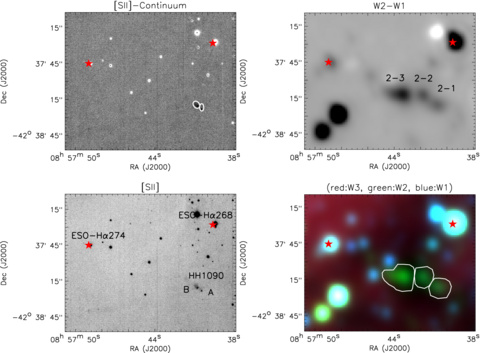}
\caption{Region of HH 1090 {\mybf and EGO 2.} {\mybf The backgrounds are continuum-subtracted [SII] image (\textit{top-left}), [SII] image (\textit{bottom-left}), difference image between WISE W2 and W1 bands (\textit{top-right}), and three-color image constructed with WISE W1 (red), W2 (green), and W3 (red) images (\textit{bottom-right}). The YSOs, YSO candidates, and protostellar core candidates are marked with red stars. The white ellipses mark the elliptic apertures that are used to measure the fluxes of HH object features while the white polygons show the polygonal apertures that are used to measure the flux densities of EGO features.} \label{fig:nhh1}}
\end{figure}

\begin{figure}
\epsscale{1.0}
\plotone{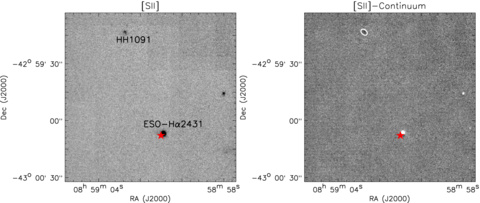}
\caption{{\mybf Region of HH 1091. The backgrounds are [SII] image (\textit{left}) and continuum-subtracted [SII] image (\textit{right}). The YSOs, YSO candidates, and protostellar core candidates are marked with red stars. The white ellipse in the right panel shows the elliptic aperture that is used to measure the flux of the HH object feature in the region.}\label{fig:nhh2}}
\end{figure}

\begin{figure}
\epsscale{1.0}
\plotone{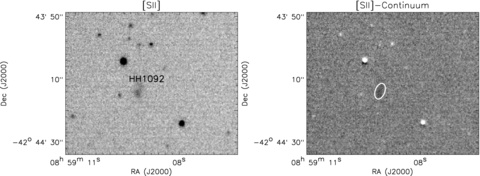}
\caption{The same as in Fig.~\ref{fig:nhh2}, but for the region of HH 1092.\label{fig:nhh3}}
\end{figure}

\begin{figure}
\epsscale{1.0}
\plotone{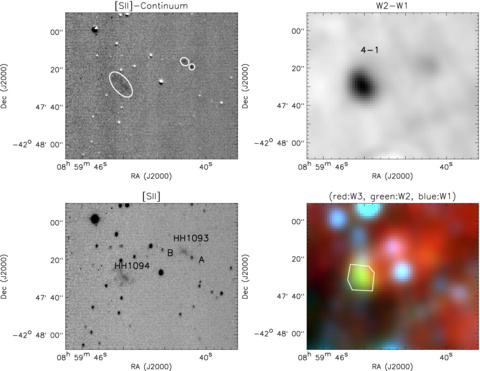}
\caption{The same as in Fig.~\ref{fig:nhh1}, but for the region of {\mybftwo HH} 1093-1094 and EGO 4.\label{fig:nhh45}}
\end{figure}

{\mybf We calculate the outflow parameters for these 10 outflows, including the lengths and position angles of outflows. For each outflow, the maximum of the distances between the exciting source and the features of its associated HH objects and EGOs is adopted as the length of this outflow if this outflow is a unipolar outflow. For the bipolar outflow, its length is the accumulation of the lengths of two lobes. The mean position angle of the features of HH objects and EGOs relative to their exciting source is adopted as the jet position angle. Note that the jet position angles have been transformed to the range of 0\degr-180\degr. Figure~\ref{fig:dist} shows the histograms of jet lengths and jet position angles.}

{\mybf The distribution of jet lengths shows a decrease in the number of outflows with increasing length. The median length of outflows in Vela C is 0.35\,pc, with the minimum of 0.15\,pc and the maximum of 2.39\,pc. \citet{gian13} investigated the jets in Vela D using the \textit{Spitzer} IRAC data and they identified 15 jets in Vela D. The median length of jets in Vela D is about 0.26\,pc, which is shorter than the median length of our outflow sample especially if considering the fact that most of outflows in our sample are unipolar rather than bipolar. However, the projection lengths of our outflow sample are mainly calculated based on the HH objects while the lengths of jets in Vela D are calculated based on the mid-infrared emission features. The optical and mid-infrared emission from an outflow trace different parts of the outflow: given the extinction, the mid-infrared emission can trace the parts that are embedded in the cloud and close to the driving source, but HH objects {\mybftwo as optical manifestation of mass outflows} can only trace the external parts that approach or perforate the surfaces of dense clouds. This situation also can explain why most of outflows detected by us are unipolar if considering the inclinations of outflows.}

\begin{figure}
\epsscale{1.0}
\plotone{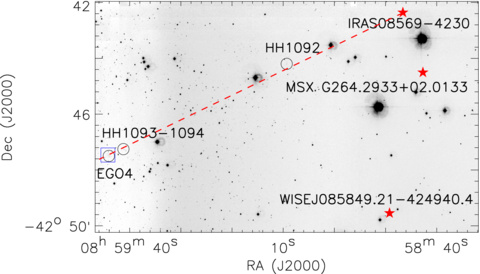}
\caption{Large view of the region of HH 1092-1094 {\mybf and EGO 4}. The background is the [SII] image. {\mybf The HH objects are marked with black circles and the EGO is labeled with blue box. The YSOs, YSO candidates, and protostellar core candidates are marked with red stars.} The {\mybf red} dashed line shows the alignment between HH 1092-1094, {\mybf EGO 4}, and the YSO IRAS 08569-4230.\label{fig:nhh345}}
\end{figure}

\begin{figure}
\epsscale{1.0}
\plotone{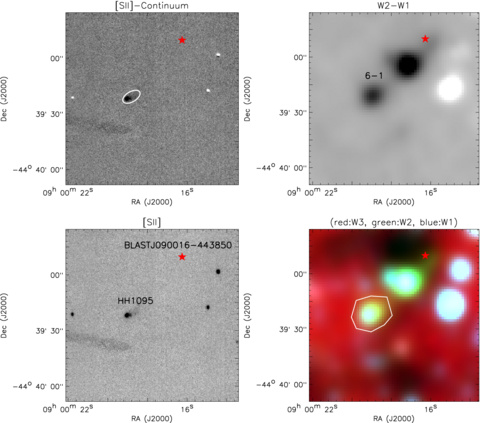}
\caption{The same as in Fig.~\ref{fig:nhh1}, but for the region of HH 1095 {\mybf and EGO 6}.\label{fig:nhh6}}
\end{figure}

\begin{figure}
\epsscale{0.8}
\plotone{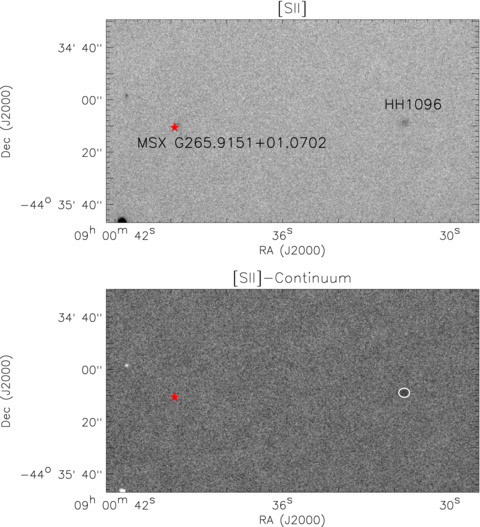}
\caption{{\mybf Region of HH 1096. The backgrounds are [SII] image (\textit{top}) and continuum-subtracted [SII] image (\textit{bottom}). The YSOs and YSO candidates are marked with red stars. The white ellipse shows the elliptic aperture that is used to measure the flux of HH object feature.}\label{fig:nhh7}}
\end{figure}

\begin{figure}
\epsscale{1.0}
\plotone{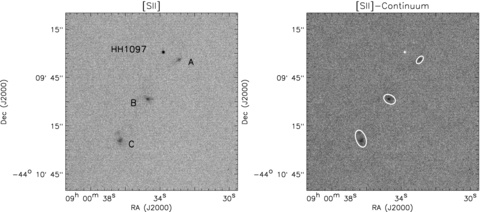}
\caption{The same as in Fig.~\ref{fig:nhh2}, but for the region of HH 1097.\label{fig:nhh8}}
\end{figure}

\begin{figure}
\epsscale{1.0}
\plotone{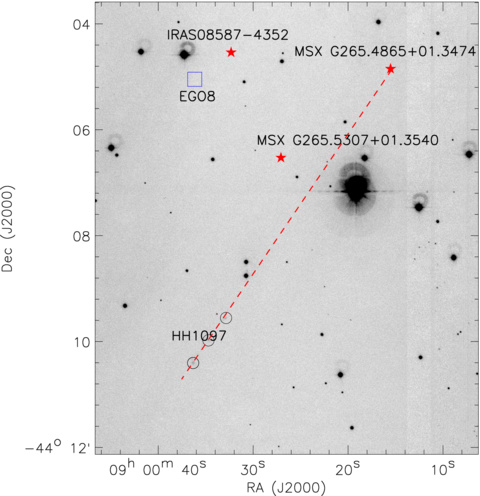}
\caption{Large view of the region of HH 1097. The {\mybf red} dashed line indicates the orientation of the alignment between HH 1097A-C. Others are the same as Fig.~\ref{fig:nhh345}.\label{fig:nhh8large}}
\end{figure}

\begin{figure}
\epsscale{1.0}
\plotone{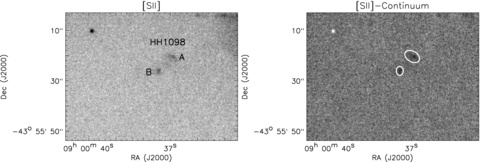}
\caption{The same as in Fig.~\ref{fig:nhh2}, but for the region of HH 1098.\label{fig:nhh9}}
\end{figure}

\begin{figure}
\epsscale{0.8}
\plotone{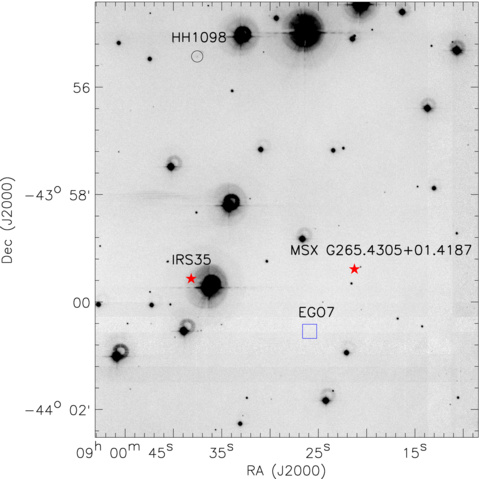}
\caption{Large view of the region of HH 1098. Others are the same as Fig.~\ref{fig:nhh345}.\label{fig:nhh9large}}
\end{figure}

\begin{figure}
\epsscale{1.0}
\plotone{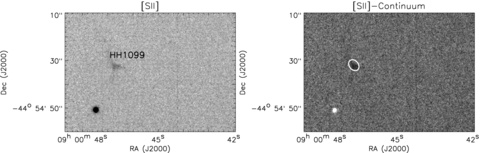}
\caption{The same as in Fig.~\ref{fig:nhh2}, but for the region of HH 1099.\label{fig:nhh10}}
\end{figure}

\begin{figure}
\epsscale{1.0}
\plotone{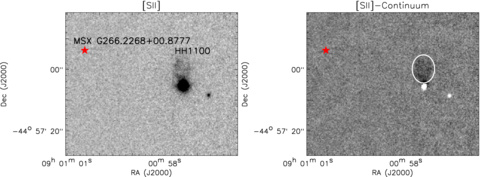}
\caption{The same as in Fig.~\ref{fig:nhh2}, but for the region of HH 1100.\label{fig:nhh11}}
\end{figure}

\begin{figure}
\epsscale{1.0}
\plotone{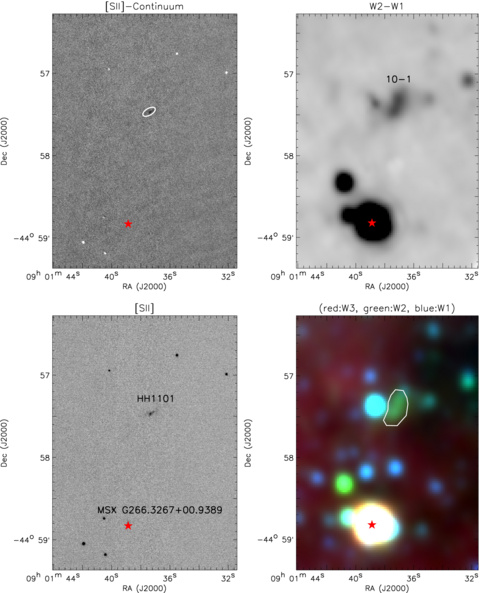}
\caption{The same as in Fig.~\ref{fig:nhh1}, but for the region of HH 1101 {\mybf and EGO 10}.\label{fig:nhh12}}
\end{figure}

\begin{figure}
\epsscale{1.0}
\plotone{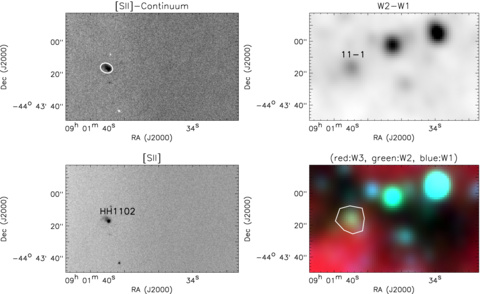}
\caption{The same as in Fig.~\ref{fig:nhh1}, but for the region of HH 1102 {\mybf and EGO 11}.\label{fig:nhh13}}
\end{figure}

\begin{figure}
\epsscale{1.0}
\plotone{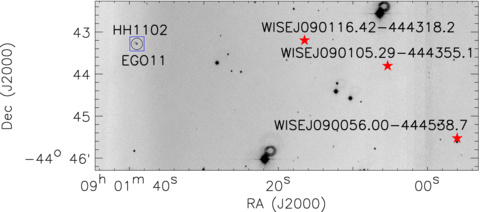}
\caption{{\mybf Large view of the region of HH 1102 and EGO 11. Others are the same as Fig.~\ref{fig:nhh345}.}\label{fig:nhh13large}}
\end{figure}

\begin{figure}
\epsscale{1.0}
\plotone{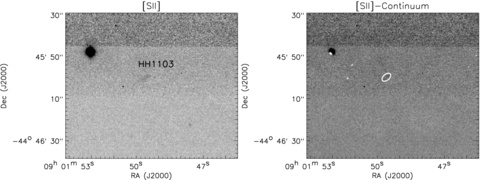}
\caption{The same as in Fig.~\ref{fig:nhh2}, but for the region of HH 1103.\label{fig:nhh14}}
\end{figure}

\begin{figure}
\epsscale{1.0}
\plotone{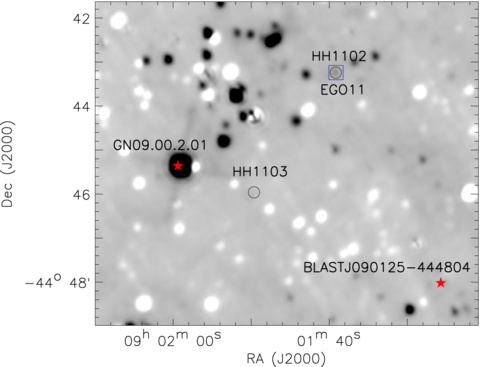}
\caption{{\mybf Large view of the region of HH 1102-1103. The background is the difference image between 3.4\,\micron~and 4.6\,\micron. Others are the same as Fig.~\ref{fig:nhh345}}.\label{fig:nhh1314large}}
\end{figure}

\begin{figure}
\epsscale{1.0}
\plotone{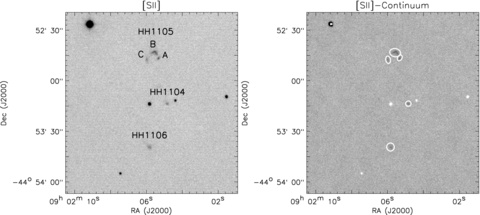}
\caption{The same as in Fig.~\ref{fig:nhh2}, but for the region of HH 1104-1106.\label{fig:nhh151617}}
\end{figure}

\begin{figure}
\epsscale{1.0}
\plotone{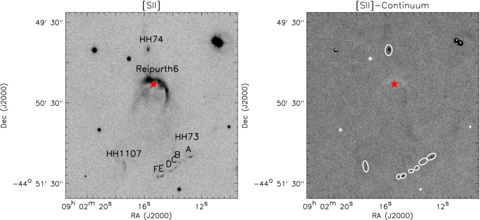}
\caption{The same as in Fig.~\ref{fig:nhh2}, but for the region of HH 1107 and HH 73-74.\label{fig:hh}}
\end{figure}

\begin{figure}
\epsscale{1.0}
\plotone{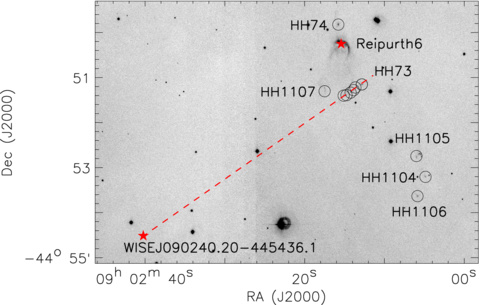}
\caption{{\mybf Large view of the region of HH 1104-1107 and HH 73. The red dashed line connecting the features of HH 73 shows the alignment between HH 73 and WISE J090240.20-445436.1. Others are the same as Fig.~\ref{fig:nhh345}.} \label{fig:hhlarge}}
\end{figure}

{\mybf The distribution of jet position angles seems to be an uniform distribution, which indicates that the outflows in Vela C are oriented randomly. Similar results have been found in several star-forming regions such as Ophiuchus \citep{zhang13}, Perseus \citep{davis08}, and Orion A \citep{davis09}. On the other hand, the study on DR21/W75 by \citet{davis07} found that the molecular hydrogen outflows, in particular from massive cores, are preferentially orthogonal to the molecular ridge, indicating a physical connection between PAs of outflows and the large-scale cloud structure. \citet{gian13} also investigated the orientation of outflows in Vela D and found that over a half of outflows have a position angle of $\sim$160\degr$\pm$15\degr~roughly parallel to the Galactic disk, which suggests that the Galactic magnetic field may be important during the early phases of the pre-main sequence evolution. However, due to the small size of our outflow sample, it is hard to do a statistical test for the distribution of jet position angles in our sample. More samples are needed to observationally confirm the conclusion that orientation of outflows has a physical connection to the large scale magnetic structure.}
{\mybf \section{Notes on individual objects}\label{sect:notes}}
{\mybf \subsection{Notes and discussion of individual HH objects and their mid-IR counterparts}}
The 20 HH objects detected in the Vela C molecular cloud are shown in Figs.~\ref{fig:nhh1}$-$\ref{fig:hhlarge}{\mybf , among which five HH objects also have mid-IR counterparts.} In this section we discuss each HH object and its possible driving sources.

Figure~\ref{fig:nhh1} shows the region of HH 1090 {\mybf and EGO 2}. HH 1090 consists of two features, HH 1090A and B. HH 1090A is a bright knot while HH 1090B is a bright knot with a diffuse patch. The distance between them is about 4\arcsec. {\mybf HH 1090 has a mid-IR counterpart, EGO 2, which exhibits as a chain of knots. EGO 2-1 corresponds to HH 1090 and EGO 2-2 and 2-3 have no optical counterparts in our [SII] image. }
There are several YSO candidates in the ambient of HH 1090. {\mybf Due to the good alignment between HH 1090, EGO 2, and ESO-H$\alpha$ 274 \citep{pre94}, we propose ESO-H$\alpha$ 274 as the possible driving source of HH 1090 and EGO 2.}

HH 1091 is a patch in Fig.~\ref{fig:nhh2}, around which there is a K5 PMS\citep{pettersson08}, ESO-H$\alpha$ 2431. The distance between ESO-H$\alpha$ 2431 and HH 1091 is about 1\arcmin, corresponding to $\sim$0.2\,pc if assuming a distance of $\sim$700\,pc for the Vela C molecular cloud.

Figure~\ref{fig:nhh3} shows the region of HH 1092, which is a faint elongated patch. Figure~\ref{fig:nhh45} shows the region of HH 1093-1094 {\mybf and EGO 4}. HH 1093 consists of two features, knot A and patch B while HH 1094 is a faint diffuse patch. {\mybf EGO 4 shows as an elongated knot in the difference image between WISE W1 and W2 band and it is the mid-IR counterpart of HH 1094.} Figure~\ref{fig:nhh345} shows the large-scale view of the region of HH 1092-1094 {\mybf and EGO 4}. There are {\mybf several} YSOs around HH 1092 within 5\arcmin. The nearest YSO to HH 1092 is a K2 PMS, ESO-H$\alpha$2430 \citep{pettersson08}, lying to the northwest about 3.6\arcmin~($\sim$0.7\,pc)~away from HH 1092 {\mybf (outside the boundary of Fig.~\ref{fig:nhh345})}.  
Note especially that there is a protostar, IRAS 08569-4230 \citep{baba06}, lying roughly in the northwest direction of the line connecting HH1092-1094. The distance between IRAS 08569-4230 and HH 1092 is about 4.5\arcmin~($\sim$1\,pc) and the distance from HH 1092 to HH 1094 is about 7.1\arcmin~($\sim$1.4\,pc). Therefore, HH 1092-1094 may belong to a parsec scale outflow which is driven by IRAS 08569-4230.

Figure~\ref{fig:nhh6} shows the region of HH 1095 {\mybf and EGO 6}. HH 1095 is a bright knot with a faint curved tail {\mybf and its mid-IR counterpart, EGO 6, is a elongated knot.} BLAST J090016-443850 (see Table.~\ref{blastproto}) 
is located about 43\arcsec~($\sim$0.15\,pc) to the northwest of HH 1095. {\mybf It seems that the tail of HH 1095 points to BLAST J090016-443850. Thus there may be physical connection between HH 1095, EGO 6, and BLAST J090016-443850.}

Figure~\ref{fig:nhh7} shows the region of HH 1096 which is a faint patch. The nearest YSO around HH 1096 is a protostar, MSX G265.9151$+$01.0702 \citep{baba06}, which is located to the east of HH 1096 at a distance of $\sim$1\farcm5 ($\sim$0.3\,pc).

Figure~\ref{fig:nhh8} shows the region of HH 1097. HH 1097 consists of three patches that align well with each other, which suggests that the driving source of HH 1097 should be located to its northwest or southeast. Figure~\ref{fig:nhh8large} shows the large-scale view of the region of HH 1097. We plotted a {\mybf red} dashed line which connects HH 1097A, B, and C and extended this line to the northwest and southeast. There are several YSO candidates lying roughly on this line, especially the protostar, MSX G265.4865$+$01.3474 \citep{baba06}. The distance from MSX G265.4865$+$01.3474 to HH 1097A is about 5.5\arcmin, corresponding to $\sim$1\,pc. Based on this perfect alignment, we suggest MSX G265.4865$+$01.3474 to be the possible driving source of HH 1097.

HH 1098 consists of two patches, A and B, shown in Fig.~\ref{fig:nhh9}. Figure~\ref{fig:nhh9large} shows the large scale view of the region of HH 1098. The nearest YSO to HH 1098 is IRS 35, a protostar identified by \citet{baba06}. Moreover, \citet{yam99b} detected a CO outflow which is driven by IRS 35. We find that HH 1098 is roughly associated with the blue lobe of this CO outflow (see Fig.~7 in \citealt{yam99b}). Therefore, we suggest IRS 35 as the likely driving source of HH 1098.

Figure~\ref{fig:nhh10} shows the region of HH 1099 which is a faint elongated patch. The nearest YSO to HH 1099 is a protostar, MSX G266.2268$+$00.8777 (outside the border of Fig.~\ref{fig:nhh10}; \citealt{baba06}), lying to the southeast about 3\farcm4 ($\sim$0.7\,pc) away from HH 1099.

Figure~\ref{fig:nhh11} shows the region of HH 1100, which exhibits a wide cavity lobe morphology that faces away from 2MASS J09005741-4457050 {\mybf (the point source below HH 1100)}. HH 1100 shows a faint diffuse structure in the [SII] image and it is different from reflection nebulae as HH 1100 also shows an obvious extended structure in the continuum-subtracted [SII] image. However, 2MASS J09005741-4457050 is not a YSO in the literature. This object is also not included in the YSO candidate sample that we have selected based on the {\mybf AllWISE} Source Catalog because of the fact that its photometry in {WISE} four bands are all contaminated by the nearby bright stars. Thus more evidence is needed to decide the nature of 2MASS J09005741-4457050.

HH 1101 is an elongated patch shown in Fig.~\ref{fig:nhh12}. {\mybf EGO 10 is the mid-IR counterpart of HH 1101 and it also shows an elongated structure.} The nearest YSO to HH 1101 is MSX G266.3267$+$00.9389, which is a protostar identified by \citet{baba06} lying to the southeast about 1\farcm4 ($\sim$0.3\,pc) away from HH 1101. 

{\mybf HH 1102 is a bright knot with a curved tail as shown in Fig.~\ref{fig:nhh13}. EGO 11 is the mid-IR counterpart of HH 1102 and shows a patch structure. Figure~\ref{fig:nhh13large} shows the large view of the region of HH 1102 and EGO 11. There are several YSO candidates to the west of HH 1102. The nearest one is WISE J090116.42-444318.2. The distance from HH 1102 to WISE J090116.42-444318.2 is $\sim$4\arcmin, corresponding to $\sim$0.8\,pc.} 

{\mybf HH 1103 is a very faint patch as shown in Fig.~\ref{fig:nhh14}.} Figure~\ref{fig:nhh1314large} shows the large-scale view of the region of HH 1102-1103. {\mybf There are several YSOs around HH 1103 and the nearest one is GN 09.00.2.01 which is a protostar candidate identified by \citet{baba06}. The distance between HH 1103 and GN 09.00.2.01 is about 1\farcm8, corresponding to $\sim$0.4\,pc.} 

Figure~\ref{fig:nhh151617} shows the region of HH 1104-1106. HH 1104 is a knot while HH 1106 is a patch. HH 1105, which consists of three features A, B, and C, shows a fragmental bow shock structure. The axis of bow shock points back to the direction of HH 1104 and 1106, which shows that there may be some physical connections among HH 1104-1106. The bow-like structure of HH 1105 indicates that the driving source is located to the south of HH 1105. However, there is no YSO or YSO candidate identified in this direction. Figure~\ref{fig:hhlarge} also shows the large-scale view of the region of HH 1104-1106. More evidence is needed to identify the driving source of HH 1104-1106.

{\mybf HH 1107 is a very faint elongated patch and is shown in Fig.~\ref{fig:hh}}. There are also two previously known HH objects in this region, HH 73 and HH 74, which are both discovered by \citet{reipurth88}. HH 73 was reported to be a chain of very faint knots by \citet{reipurth88}. Due to the higher resolution and sensitivity, this chain of knots are more obvious in our S[II] image. The chain morphology of HH 73 indicates that its driving source should be located to the northwest or southeast. We connected the knots of HH 73 with a {\mybf red} dashed line in Fig.~\ref{fig:hhlarge} and we find that {\mybf WISE J090240.20-445436.1} is located close to this line. {\mybf Although this alignment seems to be perfect, WISE J090240.20-445436.1 is a Class III source according to its spectral index (see table~\ref{ysotable}). Thus more evidence is needed to determine the driving source of HH 73. }
HH 74 was reported to be an isolated stellar-like object by \citet{reipurth88}. In our images, it also shows some faint diffuse structures around a bright knot. \citet{reipurth88} suggested that the driving source of HH 74 is also responsible for the reflection nebula, Reipurth 6 \citep{re6} which is obvious in our [SII] image. A young infrared source, HH 74 IRS, is located in the nebula, which is suggested to be the driving source of HH 74 \citep{molinari93}. HH 1107 is located near the line drawn from HH 74 to HH 74 IRS. Therefore, HH 74 IRS may be also responsible for HH 1107. 

{\mybf \subsection{Notes and discussion of individual EGOs without HH counterparts}}

\begin{figure}
\epsscale{1.0}
\plotone{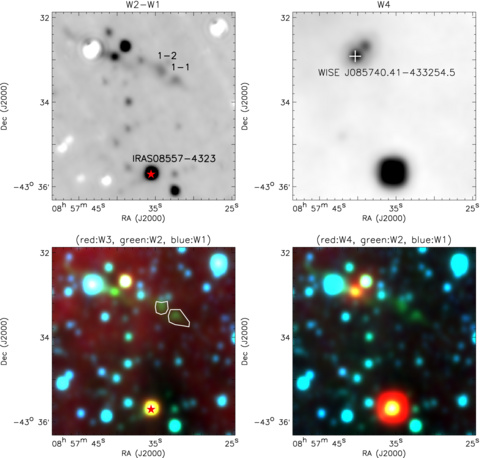}
\caption{{\mybf Region of EGO 1. The backgrounds are the difference image of WISE channel 2 (4.6\,\micron) minus channel 1 (3.4\,\micron) (\textit{top-left}), the WISE 22\micron~image (\textit{top-right}), the three-color image constructed with WISE 3.4\,\micron~(blue), 4.6\,\micron~(green), and 12.0\,\micron~(red) bands (\textit{bottom-left}), and the three-color image constructed with WISE 3.4\,\micron~(blue), 4.6\,\micron~(green), and 22\,\micron~(red) bands (\textit{bottom-right}). The YSOs and YSO candidates are marked with red stars. The white polygons show the polygonal apertures that are used to measure the fluxes of EGO features. The white plus marks the WISE source which is discussed in the text.}\label{fig:ego1}}
\end{figure}

\begin{figure}
\epsscale{1.0}
\plotone{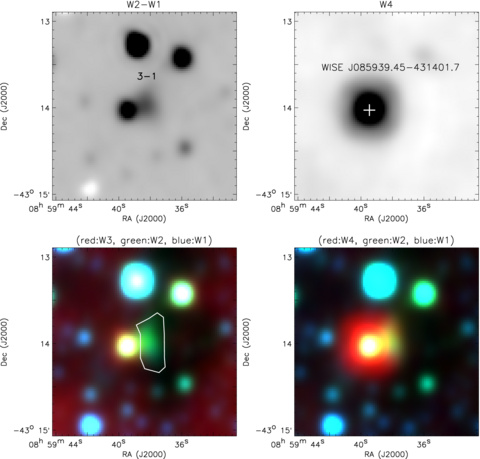}
\caption{{\mybf The same as in Fig.~\ref{fig:ego1}, but for the region of EGO 3.}\label{fig:ego3}}
\end{figure}

\begin{figure}
\epsscale{1.0}
\plotone{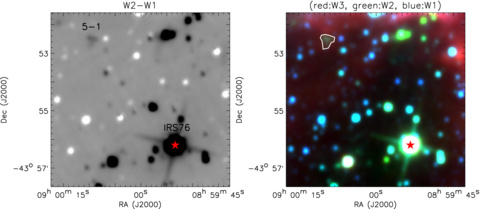}
\caption{{\mybf Region of EGO 5. The backgrounds are the difference image of WISE channel 2 (4.6\,\micron) minus channel 1 (3.4\,\micron) (\textit{left}) and the three-color image constructed with WISE 3.4\,\micron~(blue), 4.6\,\micron~(green), and 12.0\,\micron~(red) bands (\textit{right}). Others are the same as Fig.~\ref{fig:ego1}.}\label{fig:ego5}}
\end{figure}

\begin{figure}
\epsscale{1.0}
\plotone{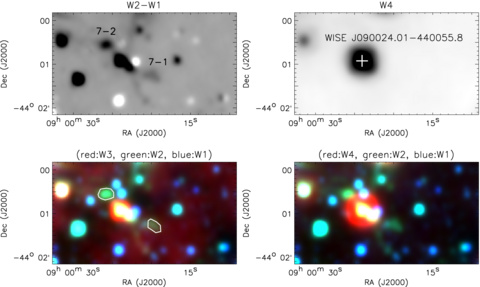}
\caption{{\mybf The same as in Fig.~\ref{fig:ego1}, but for the region of EGO 7.}\label{fig:ego7}}
\end{figure}

\begin{figure}
\epsscale{1.0}
\plotone{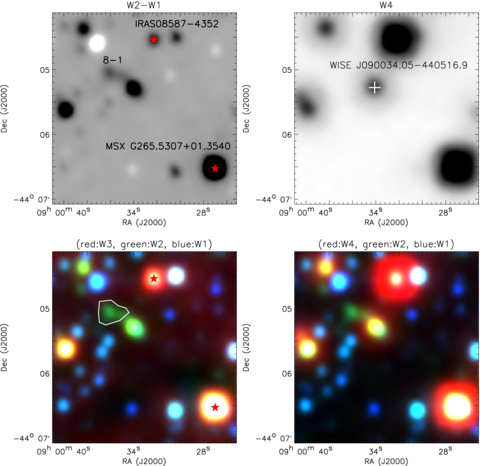}
\caption{{\mybf The same as in Fig.~\ref{fig:ego1}, but for the region of EGO 8.}\label{fig:ego8}}
\end{figure}

\begin{figure}
\epsscale{1.0}
\plotone{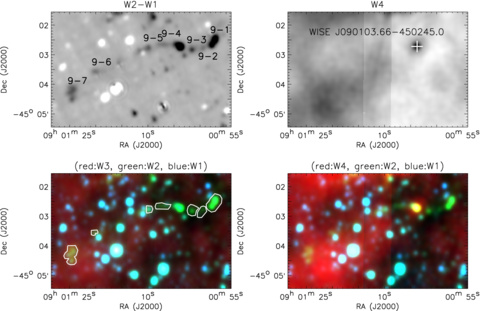}
\caption{{\mybf The same as in Fig.~\ref{fig:ego1}, but for the region of EGO 9.}\label{fig:ego9}}
\end{figure}

{\mybf We identified six EGOs without HH counterparts as new mid-infrared outflows in the Vela C molecular cloud. Figures~\ref{fig:ego1}-\ref{fig:ego9} show their images, including a difference image of WISE channel 2 (4.6\,\micron) minus channel 1 (3.4\,\micron) and a three-color image constructed with WISE 3.4\,\micron~(blue), 4.6\,\micron~(green), and 12.0\,\micron~(red) bands for each EGO. For some EGOs, we also provide a WISE 22\,\micron~image and a three-color image constructed with WISE 3.4\,\micron~(blue), 4.6\,\micron~(green), and 22\,\micron~(red) bands. In this section, we discuss each EGO and its possible driving sources.}

{\mybftwo Mid-infrared outflows are good tracers of embedded young stars. In this section we also propose five WISE sources as new YSO candidates based on the locations of EGOs and the alignments between EGOs and WISE sources. These five WISE sources are not included in our HQcatalog because their photometry in one or more WISE bands are contaminated by the artifacts (cc\_flags$\neq$`0000'). Thus they are not identified as YSO candidates in section~\ref{sect:wiseyso}. However, if we ignore the contamination and calculate their spectral indices in the wavelength range from 2\,\micron~(2MASS Ks band) to 22\,\micron~(WISE W4 band), we can obtain the values of $\alpha >$ 0, which indicates that these five sources may be protostars. These five WISE sources are marked with white pluses in Fig.~\ref{fig:ego1}-\ref{fig:ego3} and Fig.~\ref{fig:ego7}-\ref{fig:ego9} and their possible associations with the mid-infrared outflows in these regions are discussed.}

{\mybf Figure~\ref{fig:ego1} shows the region of EGO 1.  EGO 1 consists of two features, patch 1-1 and patch 1-2. The nearest YSO to EGO 1 is IRAS 08557-4323, which is a protostar candidate identified by \citet{baba06}. However, the orientation of the alignment between the two features of EGO 1 indicates that its driving source should be located to the northeast or southwest of EGO 1. In Fig.~\ref{fig:ego1}, some faint diffuse structures also can be seen to the east of EGO 1. We note that a `green' source is located to the northeast of EGO 1 in the three-color image constructed with WISE 3.4\micron~(blue), 4.6\micron~(green), and 12.0\micron~(red) bands. It corresponds to an AllWISE source, WISE J085740.41-433254.5, which is marked with a white plus in Fig.~\ref{fig:ego1}. WISE J085740.41-433254.5 is a `red' source in the three-color image constructed with WISE 3.4\micron~(blue), 4.6\micron~(green), and 22\micron~(red) bands. 
Thus we suspect that WISE J085740.41-433254.5 could be a YSO candidate and drive the EGO 1 based on the good alignment between WISE J085740.41-433254.5 and EGO 1.}

{\mybf Figure~\ref{fig:ego3} shows the image of EGO 3. EGO 3 seems to be an outflow cavity wall. The closest source to EGO 3 is WISE J085939.45-431401.7 that is marked with a white plus in Fig.~\ref{fig:ego3}. We note that WISE J085939.45-431401.7 is bright in WISE 22\micron~image and exhibit red color in the three-color image constructed with WISE 3.4\micron~(blue), 4.6\micron~(green), and 22\micron~(red) bands. We suspect that WISE J085939.45-431401.7 could be a YSO candidate and drive EGO 3.}

{\mybf Figure~\ref{fig:ego5} shows the region of EGO 5. EGO 5 exhibits a bow-like structure facing away from IRS 76. IRS 76 is a protostar candidate identified by \citet{baba06}. Thus we propose IRS 76 to be the driving source of EGO 5.}

{\mybf EGO 7 consists of two features, faint patch 7-1 and elongated knot 7-2 in Fig.~\ref{fig:ego7}. We note that a `red' source is located between the two features of EGO 7 in the three-color image constructed with WISE 3.4\micron~(blue), 4.6\micron~(green), and 12.0\micron~(red) bands, WISE J090024.01-440055.8 that is marked with a white plus in Fig.~\ref{fig:ego7}. We suspect that WISE J090024.01-440055.8 could be a YSO candidate and drive the EGO 7 bipolar outflow.}

{\mybf EGO 8 shows an NE-SW elongated structure in Fig.~\ref{fig:ego8}. The morphology of EGO 8 indicates that its driving source could be located in the northeast or southwest of EGO 8. The protostar candidate, MSX G265.5307+01.3540 \citep{baba06}, is located to the southwest of EGO 8. Thus there may be some physical connection between EGO 8 and MSX G265.5307+01.3540. We also note that a `green' source is located in the vicinity of EGO 8 and seems to have some connection with EGO 8. This `green' source corresponds to two AllWISE sources and one of them is visible in WISE 22\micron~band, WISE J090034.05-440516.9, which is marked with a white plus in Fig.~\ref{fig:ego8}. We suspect that WISE J090034.05-440516.9 could be also a YSO candidate and there may be some physical connection between WISE J090034.05-440516.9 and EGO 8.}

{\mybf Figure~\ref{fig:ego9} shows the image of EGO 9 which exhibits an S-shape. We identify seven features in EGO 9, 9-1 to 9-7. Note that the `green' source between 9-3 and 9-4 corresponds to a point source in the AllWISE source catalog, WISE J090103.66-450245.0, which is visible at 22\,\micron. We suspect that WISE J090103.66-450245.0 could be a YSO candidate and drive the EGO 9 bipolar outflow.}

\section{Summary}
We have performed a deep [SII] $\lambda\lambda$6717, 6731 wide field HH object survey toward the Vela C molecular cloud with a sky coverage of about 2 deg$^2$. Our survey covers the main part of the Vela C molecular cloud in which 13 C$^{18}$O dense clumps have been detected by \citet{yam99b}. {\mybf We also present an investigation of mid-infrared outflows in the Vela C molecular cloud using the WISE images taken from AllWISE data release.} The results are summarized in the following:

\begin{itemize}
\item[1.] In total, 20 HH objects have been detected in the Vela C molecular cloud, including 18 newly discovered HH objects, HH 1090-1107, and two previously known HH objects, HH 73-74. These 20 HH objects are mainly concentrated on the north and south region of the Vela C molecular cloud.

\item[2.] {\mybf Using the method suggested by \citet{zw09}, we identified 11 extended green objects to be mid-infrared outflows in the Vela C molecular cloud, including 6 new mid-infrared outflows which have not been detected previously at other wavelengths and 5 mid-infrared counterparts of the HH objects detected in this work.}

\item[3.] Based on the source classification scheme suggested by \citet{koenig12}, {\mybf 56} new YSO candidates are identified in Vela C using the {\mybf AllWISE} Source Catalog, including {\mybf 9 Class I sources, 6 ``Flat spectrum" sources, 29 Class II sources, and 12 Class III sources}. We also propose {\mybf 4} BLAST \citep{netterfield09} sources that are associated with the {WISE} 22\,\micron~sources to be the protostellar core candidates in the Vela C molecular cloud. 

\item[4.] Based on the morphologies of HH objects {\mybf and EGOs} and the locations of HH objects{\mybf , EGOs} and YSOs, the possible driving sources of the HH objects {\mybf and EGOs} are discussed. {\mybf We associate 12 HH objects and five EGOs with 10 YSOs and YSO candidates. The median length of these 10 outflows is 0.35\,pc and the distribution of jet position angles seems to be uniform.}

\end{itemize}
\acknowledgements
We thank the ESO/MPG 2.2m WFI group for making the observations in service mode. H. Wang acknowledges the support by NSFC grants 11173060, 11233007, and 11127903. This work is supported by the Strategic Priority Research Program ``The Emergence of Cosmological Structures" of the Chinese Academy of Sciences, Grant No. XDB09000000. This publication makes use of data products from the Wide-field Infrared Survey Explorer, which is a joint project of the University of California, Los Angeles, and the Jet Propulsion Laboratory/California Institute of Technology, funded by the National Aeronautics and Space Administration. This research was made possible through the use of the AAVSO Photometric All-Sky Survey (APASS), funded by the Robert Martin Ayers Sciences Fund. This research has also made use of the SIMBAD database, operated at CDS, Strasbourg, France. In particular, the Aladin image server is very useful in locating previously known objects on our frames. This paper is part of an ongoing collaboration between the Purple Mountain Observatory in Nanjing and the Max Planck Institute for Astronomy in Heidelberg. 

\clearpage

\begin{deluxetable}{p{2.5cm}cclccc}
\rotate
\tabletypesize{\tiny}
\tablecolumns{7}
\tablewidth{0pc}
\tablecaption{Newly Discovered and Previously Known Herbig-Haro Objects in Vela C \label{tab1}}
\tablehead{
\colhead{}&
\colhead{$\alpha$\tablenotemark{a}}&
\colhead{$\delta$\tablenotemark{a}}&
\colhead{}&
\colhead{Flux$_{peak}$\tablenotemark{b}}&
\colhead{SNR$_{peak}$\tablenotemark{c}}&
\colhead{Flux$_{total}$\tablenotemark{d}}\\
\colhead{Object} &
\colhead{(J2000.0)} &
\colhead{(J2000.0)} &
\colhead{Comments}&
\colhead{(10$^{-19}$ W m$^{-2}$ arcsec$^{-2}$)}&
\colhead{}&
\colhead{(10$^{-19}$ W m$^{-2}$)}
}
\startdata
HH1090A\dotfill& 08 57 40.4  &-42 38 23&knot&0.83&10.3&2.42$\pm$0.12\\
HH1090B\dotfill& 08 57 40.8  &-42 38 21&knot and diffuse patch         &1.14&14.2&5.41$\pm$0.16\\
HH1091\dotfill& 08 59 02.7  &-42 59 13&patch                          &0.49&4.8&1.14$\pm$0.10\\
HH1092\dotfill& 08 59 09.2  &-42 44 14&faint elongated patch          &0.30&4.2&1.21$\pm$0.09\\
HH1093A\dotfill& 08 59 40.7  &-42 47 19&knot                           &0.86&8.5&1.45$\pm$0.10\\
HH1093B\dotfill& 08 59 41.1  &-42 47 16&patch                          &0.47&4.6&2.07$\pm$0.14\\
HH1094\dotfill& 08 59 44.0  &-42 47 30&faint diffuse patch            &0.67&6.1&13.55$\pm$0.40\\
HH1095\dotfill& 09 00 19.1  &-44 39 22&bright knot with a curved tail &3.21&30.0&12.28$\pm$0.27\\
HH1096\dotfill& 09 00 31.7  &-44 35 09&faint patch                    &0.36&3.8&1.26$\pm$0.11\\
HH1097A\dotfill& 09 00 32.8  &-44 09 34&patch                          &0.46&5.4&1.54$\pm$0.11\\
HH1097B\dotfill& 09 00 34.7  &-44 09 59&patch                          &0.64&7.6&3.61$\pm$0.17\\
HH1097C\dotfill& 09 00 36.3  &-44 10 25&patch                          &0.54&6.4&5.54$\pm$0.22\\
HH1098A\dotfill& 09 00 36.9  &-43 55 20&patch                          &0.42&5.6&1.85$\pm$0.12\\
HH1098B\dotfill& 09 00 37.4  &-43 55 26&patch                          &0.37&5.0&0.94$\pm$0.08\\
HH1099\dotfill& 09 00 46.6  &-44 54 32&faint patch                    &0.42&4.3&1.65$\pm$0.13\\
HH1100\dotfill& 09 00 57.4  &-44 57 04&wide cavity lobe     &1.89&21.2&9.12$\pm$0.24\\
HH1101\dotfill& 09 01 37.4  &-44 57 28&elongated patch                &0.69&6.4&5.47$\pm$0.24\\
HH1102\dotfill& 09 01 38.9  &-44 43 17&bright knot with a tail              &1.73&18.1&10.13$\pm$0.22\\
HH1103\dotfill& 09 01 49.5  &-44 46 00&faint patch                    &0.57&3.4&1.40$\pm$0.19\\
HH1104\dotfill& 09 02 04.8  &-44 53 14&knot                           &0.64&7.0&1.17$\pm$0.09\\
HH1105A\dotfill& 09 02 05.3  &-44 52 47&knot, part of the bow shock    &0.73&8.1&1.39$\pm$0.09\\
HH1105B\dotfill& 09 02 05.5  &-44 52 43&bow shock                      &0.70&7.7&4.36$\pm$0.17\\
HH1106\dotfill& 09 02 05.8  &-44 53 40&patch                          &0.45&4.9&1.93$\pm$0.14\\
HH1105C\dotfill& 09 02 06.0  &-44 52 46&patch, part of the bow shock   &0.44&4.8&1.51$\pm$0.10\\
HH73A\dotfill& 09 02 12.8  &-44 51 11&knot                           &0.63&7.0&2.92$\pm$0.16\\
HH73B\dotfill& 09 02 13.6  &-44 51 15&knot                           &0.43&4.7&1.81$\pm$0.15\\
HH73C\dotfill& 09 02 13.8  &-44 51 18&knot                           &0.40&4.4&1.05$\pm$0.10\\
HH73D\dotfill& 09 02 14.2  &-44 51 21&knot                           &0.44&4.8&0.77$\pm$0.09\\
HH73E\dotfill& 09 02 14.7  &-44 51 25&knot                           &0.56&6.2&1.73$\pm$0.11\\
HH73F\dotfill& 09 02 15.0  &-44 51 25&knot                           &0.62&6.9&1.25$\pm$0.10\\
HH74\dotfill& 09 02 15.7  &-44 49 51&bright knot and diffuse patch  &0.89&9.8&4.81$\pm$0.19\\
HH1107\dotfill& 09 02 17.4  &-44 51 19&faint patch                    &0.43&4.5&2.47$\pm$0.17\\
\enddata
\tablecomments{Units of right ascension are hours, minutes, and seconds, and units of declination are degrees, arcminutes, and arcseconds.}
\tablenotetext{a}{Peak position on the continuum-subtracted [SII] image.}
\tablenotetext{b}{Surface brightness of peak in HH feature.}
\tablenotetext{c}{Signal to noise ratio of peak in HH feature.}
\tablenotetext{d}{Total flux of HH feature estimated using elliptical aperture.}
\end{deluxetable}

\clearpage

\begin{deluxetable}{cclccc}
\tabletypesize{\scriptsize}
\tablecolumns{6}
\tablewidth{0pc}
\tablecaption{Extended green objects in Vela C \label{egotable}}
\tablehead{
\colhead{}&
\colhead{$\alpha$\tablenotemark{a}}&
\colhead{$\delta$\tablenotemark{a}}&
\colhead{}&
\colhead{Number of}&
\colhead{HH}\\
\colhead{ID}&
\colhead{(J2000)}&
\colhead{(J2000)}&
\colhead{EGO}&
\colhead{features}&
\colhead{counterparts}}
\startdata
1& 08 57 33.4  &-43 33 19&J085733.4-433319&2&\ldots\\
2& 08 57 42.0  &-42 38 15&J085742.0-423815&3&HH1090\\
3& 08 59 38.5  &-43 13 59&J085938.5-431359&1&\ldots\\
4& 08 59 44.2  &-42 47 28&J085944.2-424728&1&HH1094\\
5& 09 00 09.6  &-43 52 29&J090009.6-435229&1&\ldots\\
6& 09 00 19.0  &-44 39 21&J090019.0-443921&1&HH1095\\
7& 09 00 23.1  &-44 00 55&J090023.1-440055&2&\ldots\\
8& 09 00 36.0  &-44 05 03&J090036.0-440503&1&\ldots\\
9& 09 01 08.2  &-45 03 03&J090108.2-450303&7&\ldots\\
10& 09 01 37.2  &-44 57 26&J090137.2-445726&1&HH1101\\
11& 09 01 38.9  &-44 43 17&J090138.9-444317&1&HH1102\\
\enddata
\tablecomments{Units of right ascension are hours, minutes, and seconds, and units of declination are degrees, arcminutes, and arcseconds.}
\tablenotetext{a}{The mean coordinate of all features in the EGO.}
\end{deluxetable}

\clearpage

\begin{deluxetable}{cclccccc}
\rotate
\tabletypesize{\scriptsize}
\tablecolumns{8}
\tablewidth{0pc}
\tablecaption{Photometry of EGO features in Vela C \label{egophoto}}
\tablehead{
\colhead{ID\tablenotemark{a}}&
\colhead{$\alpha$\tablenotemark{b}}&
\colhead{$\delta$\tablenotemark{b}}&
\colhead{Size\tablenotemark{c}}&
\colhead{$F(3.4)\pm\Delta F(3.4)$\tablenotemark{d}}&
\colhead{$F(4.6)\pm\Delta F(4.6)$\tablenotemark{d}}&
\colhead{$F(12)\pm\Delta F(12)$\tablenotemark{d}}\\
\colhead{of features}&
\colhead{(J2000)}&
\colhead{(J2000)}&
\colhead{(arcsec$^2$)}&
\colhead{(mJy)}&
\colhead{(mJy)}&
\colhead{(mJy)}}
\startdata
1-1& 08 57 32.5  &-43 33 28&512     &0.62$\pm$0.04&1.28$\pm$0.06&1.59$\pm$0.06\\
1-2& 08 57 34.3  &-43 33 11&272     &0.59$\pm$0.03&1.15$\pm$0.06&1.87$\pm$0.06\\
2-1& 08 57 40.7  &-42 38 20&251     &0.56$\pm$0.04&1.50$\pm$0.06&1.72$\pm$0.13\\
2-2& 08 57 41.8  &-42 38 12&234     &0.34$\pm$0.03&1.65$\pm$0.07&\ldots\\
2-3& 08 57 43.3  &-42 38 12&449     &0.41$\pm$0.04&3.82$\pm$0.10&\ldots\\
3-1& 08 59 38.5  &-43 13 59&610     &1.35$\pm$0.05&3.39$\pm$0.10&0.25$\pm$0.04\\
4-1& 08 59 44.2  &-42 47 28&166     &0.34$\pm$0.03&0.69$\pm$0.04&0.41$\pm$0.03\\
5-1& 09 00 09.6  &-43 52 29&587     &1.80$\pm$0.06&3.21$\pm$0.09&19.24$\pm$0.20\\
6-1& 09 00 19.0  &-44 39 21&325     &1.28$\pm$0.05&1.90$\pm$0.07&2.84$\pm$0.08\\
7-1& 09 00 20.4  &-44 01 16&230     &0.34$\pm$0.03&1.20$\pm$0.06&1.95$\pm$0.07\\
7-2& 09 00 25.9  &-44 00 34&274     &0.83$\pm$0.04&3.41$\pm$0.10&1.70$\pm$0.07\\
8-1& 09 00 36.0  &-44 05 03&444     &0.36$\pm$0.03&2.20$\pm$0.08&1.17$\pm$0.06\\
9-1& 09 00 56.9  &-45 02 31&771     &0.78$\pm$0.04&5.15$\pm$0.12&\ldots\\
9-2& 09 00 59.2  &-45 02 48&302     &\ldots&0.47$\pm$0.04&\ldots\\
9-3& 09 01 01.4  &-45 02 51&370     &\ldots&0.92$\pm$0.05&\ldots\\
9-4& 09 01 06.3  &-45 02 38&321     &0.18$\pm$0.02&0.82$\pm$0.05&\ldots\\
9-5& 09 01 09.7  &-45 02 46&168     &0.69$\pm$0.04&0.98$\pm$0.05&0.68$\pm$0.05\\
9-6& 09 01 19.7  &-45 03 35&117     &0.22$\pm$0.02&0.40$\pm$0.03&1.42$\pm$0.06\\
9-7& 09 01 24.3  &-45 04 12&760     &1.80$\pm$0.06&3.31$\pm$0.09&19.51$\pm$0.21\\
10-1& 09 01 37.2  &-44 57 26&328     &0.65$\pm$0.04&1.73$\pm$0.07&0.80$\pm$0.07\\
11-1& 09 01 38.9  &-44 43 17&242     &0.35$\pm$0.03&0.71$\pm$0.04&2.10$\pm$0.07\\
\enddata
\tablecomments{Units of right ascension are hours, minutes, and seconds, and units of declination are degrees, arcminutes, and arcseconds.}
\tablenotetext{a}{The number in front of the dash corresponds to the ID in table~\ref{egotable}.}
\tablenotetext{b}{Peak position on the 4.6\,\micron~image.}
\tablenotetext{c}{Area of polygon that is defined visually based on the morphology and surface brightness distribution of EGO feature at 4.6\,\micron.}
\tablenotetext{d}{Flux estimated using polygonal aperture.}
\end{deluxetable}

\clearpage

\begin{deluxetable}{lllcllllllllll}
\rotate
\centering
\tabletypesize{\tiny}
\tablewidth{0pt}
\setlength{\tabcolsep}{0.1cm}
\tablecaption{New YSO candidates in Vela C identified with ALLWISE Source Catalog.\label{ysotable}}
\tablehead{
\colhead{}&
\colhead{}&
\colhead{}&
\colhead{Spectral}&
\multicolumn{3}{c}{2MASS}&
\multicolumn{4}{c}{WISE}&
\multicolumn{3}{c}{BLAST}\\
\cline{5-7}
\cline{8-11}
\cline{12-14}
\colhead{WISE}&
\colhead{$\alpha$}&
\colhead{$\delta$}&
\colhead{index}&
\colhead{J}&
\colhead{H}&
\colhead{Ks}&
\colhead{W1}&
\colhead{W2}&
\colhead{W3}&
\colhead{W4}&
\colhead{F250}&
\colhead{F350}&
\colhead{F500}\\
\colhead{}&
\colhead{(J2000)}&
\colhead{(J2000)}&
\colhead{($\alpha$)}&
\colhead{(mag)}&
\colhead{(mag)}&
\colhead{(mag)}&
\colhead{(mag)}&
\colhead{(mag)}&
\colhead{(mag)}&
\colhead{(mag)}&
\colhead{(Jy)}&
\colhead{(Jy)}&
\colhead{(Jy)}
}
\startdata
J085401.63-423324.8& 08 54 01.6  &-42 33 25&-0.69&\ldots&\ldots&\ldots&13.91$\pm$0.03&13.32$\pm$0.03&11.20$\pm$0.20&8.03$\pm$0.25&\ldots&\ldots&\ldots\\                                                
J085402.36-422110.6& 08 54 02.4  &-42 21 11&-1.40&15.02$\pm$0.05&13.82$\pm$0.04&13.23$\pm$0.04&12.67$\pm$0.03&12.11$\pm$0.03&10.43$\pm$0.09&7.99(u)&\ldots&\ldots&\ldots\\                              
J085405.75-423448.8& 08 54 05.8  &-42 34 49&-0.86&15.92$\pm$0.07&14.85$\pm$0.08&14.78$\pm$0.10&13.79$\pm$0.04&13.34$\pm$0.04&10.95$\pm$0.14&8.17$\pm$0.41&\ldots&\ldots&\ldots\\                        
J085407.90-423536.9& 08 54 07.9  &-42 35 37&-0.55&15.42$\pm$0.06&13.93$\pm$0.06&13.27$\pm$0.05&12.32$\pm$0.02&11.53$\pm$0.02&9.28$\pm$0.04&6.88$\pm$0.08&\ldots&\ldots&\ldots\\                         
J085409.80-451514.6& 08 54 09.8  &-45 15 15&-0.41&13.72$\pm$0.04&12.81$\pm$0.04&11.86$\pm$0.03&10.77$\pm$0.02&10.08$\pm$0.02&7.38$\pm$0.02&5.67$\pm$0.04&\ldots&\ldots&\ldots\\                         
J085410.36-430214.4& 08 54 10.4  &-43 02 14&-1.21&13.76$\pm$0.03&13.01$\pm$0.03&12.46$\pm$0.03&11.66$\pm$0.02&11.11$\pm$0.02&9.97$\pm$0.10&8.72$\pm$0.40&\ldots&\ldots&\ldots\\                         
J085413.28-435923.0& 08 54 13.3  &-43 59 23&0.45&13.90$\pm$0.03&13.38$\pm$0.04&12.82$\pm$0.03&11.16$\pm$0.02&9.96$\pm$0.02&6.47$\pm$0.01&4.65$\pm$0.02&\ldots&\ldots&\ldots\\                           
J085506.79-440739.3& 08 55 06.8  &-44 07 39&-0.59&16.49$\pm$0.14&15.11$\pm$0.07&14.19$\pm$0.07&13.20$\pm$0.03&12.54$\pm$0.03&9.96$\pm$0.07&8.32$\pm$0.28&\ldots&\ldots&\ldots\\                         
J085512.11-435826.0& 08 55 12.1  &-43 58 26&-0.99&16.36$\pm$0.11&14.98$\pm$0.07&13.75$\pm$0.05&12.78$\pm$0.02&12.23$\pm$0.02&10.45$\pm$0.12&7.86(u)&\ldots&\ldots&\ldots\\                              
J085513.14-423715.1& 08 55 13.1  &-42 37 15&-1.25&16.17$\pm$0.10&14.42$\pm$0.05&13.55$\pm$0.05&12.99$\pm$0.03&12.32$\pm$0.02&10.66$\pm$0.10&8.47$\pm$0.28&\ldots&\ldots&\ldots\\                        
J085516.00-424549.4& 08 55 16.0  &-42 45 49&-1.01&15.80$\pm$0.09&14.76$\pm$0.09&14.07$\pm$0.06&13.10$\pm$0.04&12.53$\pm$0.04&10.94$\pm$0.14&8.20(u)&\ldots&\ldots&\ldots\\                              
J085524.71-440343.4& 08 55 24.7  &-44 03 43&-2.01&11.82$\pm$0.02&11.10$\pm$0.03&10.72$\pm$0.02&10.40$\pm$0.02&10.36$\pm$0.02&10.09$\pm$0.08&6.67$\pm$0.07&\ldots&\ldots&\ldots\\                        
J085528.41-434625.3& 08 55 28.4  &-43 46 25&-0.87&16.32$\pm$0.14&15.02$\pm$0.08&14.65$\pm$0.10&13.67$\pm$0.03&13.07$\pm$0.03&10.89$\pm$0.11&8.87(u)&\ldots&\ldots&\ldots\\                              
J085534.25-432655.8& 08 55 34.3  &-43 26 56&0.18&14.41$\pm$0.04&13.18$\pm$0.03&12.38$\pm$0.03&11.41$\pm$0.02&10.27$\pm$0.02&7.13$\pm$0.02&4.16$\pm$0.03&\ldots&\ldots&\ldots\\                          
J085535.41-451013.5& 08 55 35.4  &-45 10 14&1.38&\ldots&\ldots&\ldots&14.67$\pm$0.04&13.77$\pm$0.04&9.28$\pm$0.04&5.86$\pm$0.04&\ldots&\ldots&\ldots\\                                                  
J085539.72-432554.4& 08 55 39.7  &-43 25 54&-0.51&15.98$\pm$0.11&14.70$\pm$0.06&13.98$\pm$0.08&13.47$\pm$0.03&12.79$\pm$0.03&10.04$\pm$0.05&7.88$\pm$0.17&\ldots&\ldots&\ldots\\                        
J085539.93-420144.6& 08 55 39.9  &-42 01 45&-2.27&11.11$\pm$0.02&9.98$\pm$0.02&9.54$\pm$0.02&9.28$\pm$0.02&9.30$\pm$0.02&8.80$\pm$0.06&6.80$\pm$0.08&\ldots&\ldots&\ldots\\                             
J085551.72-440628.6& 08 55 51.7  &-44 06 29&-0.98&13.75$\pm$0.03&12.85$\pm$0.02&12.23$\pm$0.03&11.43$\pm$0.02&10.94$\pm$0.02&9.23$\pm$0.04&6.37$\pm$0.06&\ldots&\ldots&\ldots\\                         
J085551.85-435622.2& 08 55 51.9  &-43 56 22&-1.31&16.17$\pm$0.11&14.86$\pm$0.09&14.22$\pm$0.09&13.42$\pm$0.03&12.95$\pm$0.03&11.19(u)&8.22(u)&\ldots&\ldots&\ldots\\                                    
J085556.52-441442.3& 08 55 56.5  &-44 14 42&-1.62&14.77$\pm$0.04&13.76$\pm$0.04&13.25$\pm$0.04&12.91$\pm$0.02&12.46$\pm$0.02&10.47$\pm$0.08&8.90$\pm$0.44&\ldots&\ldots&\ldots\\                        
J085556.74-452520.6& 08 55 56.7  &-45 25 21&-0.77&\ldots&\ldots&\ldots&13.98$\pm$0.03&13.38$\pm$0.03&11.00$\pm$0.12&8.69(u)&\ldots&\ldots&\ldots\\                                                      
J085602.09-434735.7& 08 56 02.1  &-43 47 36&-0.94&15.31$\pm$0.06&13.93$\pm$0.05&13.19$\pm$0.04&12.42$\pm$0.02&11.94$\pm$0.02&9.58$\pm$0.07&7.24$\pm$0.12&\ldots&\ldots&\ldots\\                         
J085616.80-423313.8& 08 56 16.8  &-42 33 14&0.00&16.25$\pm$0.13&13.67$\pm$0.03&11.93$\pm$0.03&11.08$\pm$0.02&9.55$\pm$0.02&6.90$\pm$0.02&4.04$\pm$0.03&\ldots&\ldots&\ldots\\                           
J085618.06-423431.6& 08 56 18.1  &-42 34 32&-0.25&16.74$\pm$0.19&15.26$\pm$0.11&14.42$\pm$0.08&13.47$\pm$0.03&12.56$\pm$0.02&10.21$\pm$0.07&7.18$\pm$0.10&\ldots&\ldots&\ldots\\                        
J085623.69-435236.5& 08 56 23.7  &-43 52 37&-1.19&16.30$\pm$0.15&14.72$\pm$0.07&14.15$\pm$0.08&13.50$\pm$0.03&13.09$\pm$0.03&10.73$\pm$0.12&8.60$\pm$0.45&\ldots&\ldots&\ldots\\                        
J085625.83-451444.9& 08 56 25.8  &-45 14 45&-0.49&13.79$\pm$0.03&12.51$\pm$0.03&11.39$\pm$0.02&10.44$\pm$0.02&9.67$\pm$0.02&7.10$\pm$0.02&5.18$\pm$0.03&\ldots&\ldots&\ldots\\                          
J085629.39-451846.3& 08 56 29.4  &-45 18 46&-2.13&12.33$\pm$0.02&11.85$\pm$0.03&11.57$\pm$0.02&11.44$\pm$0.02&11.38$\pm$0.02&9.75$\pm$0.07&7.05$\pm$0.10&\ldots&\ldots&\ldots\\                         
J085638.40-423714.9& 08 56 38.4  &-42 37 15&-1.17&14.23$\pm$0.03&13.21$\pm$0.03&12.83$\pm$0.03&12.30$\pm$0.02&11.87$\pm$0.02&9.47$\pm$0.06&7.26$\pm$0.09&\ldots&\ldots&\ldots\\                         
J085712.07-421935.3& 08 57 12.1  &-42 19 35&-2.29&11.49$\pm$0.02&10.11$\pm$0.02&9.57$\pm$0.02&9.30$\pm$0.02&9.38$\pm$0.02&8.69$\pm$0.02&6.87$\pm$0.07&\ldots&\ldots&\ldots\\                            
J085758.18-424630.5& 08 57 58.2  &-42 46 31&0.30&16.91$\pm$0.17&15.55$\pm$0.12&13.99$\pm$0.07&11.27$\pm$0.02&9.94$\pm$0.02&7.33$\pm$0.04&4.86$\pm$0.04&6.5$\pm$1.3&4.9$\pm$1.0&2.4$\pm$0.7\\            
J085817.33-424913.2& 08 58 17.3  &-42 49 13&-0.55&16.64$\pm$0.14&14.10$\pm$0.04&12.77$\pm$0.03&11.77$\pm$0.02&10.83$\pm$0.02&9.40$\pm$0.11&8.38$\pm$0.48&\ldots&\ldots&\ldots\\                         
J085818.97-452245.9& 08 58 19.0  &-45 22 46&-2.27&11.23$\pm$0.02&10.69$\pm$0.02&10.40$\pm$0.02&10.21$\pm$0.02&10.13$\pm$0.02&9.57$\pm$0.03&6.99$\pm$0.08&\ldots&\ldots&\ldots\\                         
J085827.43-442915.6& 08 58 27.4  &-44 29 16&-1.64&11.21$\pm$0.02&10.74$\pm$0.02&10.54$\pm$0.02&10.28$\pm$0.02&10.16$\pm$0.02&9.03$\pm$0.03&5.38$\pm$0.04&\ldots&\ldots&\ldots\\                         
J085829.01-443412.1& 08 58 29.0  &-44 34 12&-2.46&11.19$\pm$0.02&10.76$\pm$0.02&10.56$\pm$0.02&10.49$\pm$0.02&10.47$\pm$0.02&10.14$\pm$0.07&7.10$\pm$0.10&\ldots&\ldots&\ldots\\                        
J085849.21-424940.4& 08 58 49.2  &-42 49 40&-1.41&15.00$\pm$0.05&14.31$\pm$0.05&14.06$\pm$0.06&13.66$\pm$0.03&13.23$\pm$0.03&11.00$\pm$0.16&8.13$\pm$0.26&\ldots&\ldots&\ldots\\                        
J085912.89-423801.6& 08 59 12.9  &-42 38 02&-1.33&14.25$\pm$0.03&13.45$\pm$0.03&13.20$\pm$0.04&12.76$\pm$0.02&12.34$\pm$0.02&10.10$\pm$0.06&8.30$\pm$0.28&\ldots&\ldots&\ldots\\                        
J085917.70-442609.0& 08 59 17.7  &-44 26 09&-0.63&16.38$\pm$0.12&15.07$\pm$0.07&14.22$\pm$0.09&13.60$\pm$0.03&12.75$\pm$0.03&10.29$\pm$0.12&7.45$\pm$0.36&\ldots&\ldots&\ldots\\                        
J085918.05-442615.4& 08 59 18.1  &-44 26 15&1.78&\ldots&\ldots&\ldots&14.62$\pm$0.04&12.27$\pm$0.03&8.41$\pm$0.03&4.24$\pm$0.03&\ldots&\ldots&\ldots\\                                                  
J085921.32-430041.5& 08 59 21.3  &-43 00 42&-0.72&14.48$\pm$0.04&13.54$\pm$0.04&13.07$\pm$0.04&12.29$\pm$0.03&11.61$\pm$0.02&9.03$\pm$0.05&7.08$\pm$0.16&\ldots&\ldots&\ldots\\                         
J085922.13-443329.8& 08 59 22.1  &-44 33 30&1.68&\ldots&\ldots&\ldots&15.54$\pm$0.04&14.02$\pm$0.04&12.07(u)&6.06$\pm$0.05&4.2$\pm$0.7&4.6$\pm$0.7&2.7$\pm$0.5\\                                        
J085929.62-443123.4& 08 59 29.6  &-44 31 23&0.18&16.96$\pm$0.20&14.40$\pm$0.05&12.60$\pm$0.03&11.56$\pm$0.02&10.10$\pm$0.02&7.26$\pm$0.02&4.46$\pm$0.03&16.6$\pm$2.4&16.9$\pm$2.4&11.4$\pm$1.8\\        
J085939.37-443648.8& 08 59 39.4  &-44 36 49&0.14&\ldots&\ldots&\ldots&11.94$\pm$0.02&10.17$\pm$0.02&7.85$\pm$0.03&5.11$\pm$0.03&\ldots&\ldots&\ldots\\                                                  
J090051.90-442618.4& 09 00 51.9  &-44 26 18&-1.45&14.64$\pm$0.04&13.11$\pm$0.03&12.41$\pm$0.03&11.95$\pm$0.02&11.27$\pm$0.02&9.81$\pm$0.10&8.40$\pm$0.37&4.9$\pm$0.7&5.8$\pm$0.8&4.0$\pm$0.6\\          
J090056.00-444538.7& 09 00 56.0  &-44 45 39&-0.50&13.25$\pm$0.02&11.95$\pm$0.03&11.09$\pm$0.03&10.10$\pm$0.02&9.28$\pm$0.02&7.01$\pm$0.02&4.69$\pm$0.03&\ldots&\ldots&\ldots\\                          
J090105.29-444355.1& 09 01 05.3  &-44 43 55&1.30&\ldots&\ldots&\ldots&15.90$\pm$0.06&14.68$\pm$0.05&11.19(u)&7.22$\pm$0.11&\ldots&\ldots&\ldots\\                                                       
J090116.42-444318.2& 09 01 16.4  &-44 43 18&1.33&\ldots&\ldots&\ldots&15.48$\pm$0.04&13.42$\pm$0.03&11.73(u)&6.44$\pm$0.06&\ldots&\ldots&\ldots\\                                                       
J090149.42-443524.3& 09 01 49.4  &-44 35 24&-0.24&\ldots&\ldots&\ldots&13.95$\pm$0.03&13.20$\pm$0.03&10.49$\pm$0.09&7.87$\pm$0.17&\ldots&\ldots&\ldots\\                                                
J090208.62-443457.2& 09 02 08.6  &-44 34 57&1.69&\ldots&\ldots&\ldots&15.58$\pm$0.04&11.59$\pm$0.02&8.96$\pm$0.05&4.17$\pm$0.03&5.6$\pm$0.8&4.8$\pm$0.7&3.1$\pm$0.5\\                                   
J090212.76-444103.3& 09 02 12.8  &-44 41 03&-1.04&14.65$\pm$0.04&13.51$\pm$0.02&12.81$\pm$0.02&12.06$\pm$0.02&11.35$\pm$0.02&9.81$\pm$0.18&7.92$\pm$0.17&\ldots&\ldots&\ldots\\                         
J090216.86-443032.2& 09 02 16.9  &-44 30 32&1.68&\ldots&\ldots&\ldots&14.32$\pm$0.10&13.74$\pm$0.04&8.87$\pm$0.03&5.57$\pm$0.04&\ldots&\ldots&\ldots\\                                                  
J090218.63-441000.3& 09 02 18.6  &-44 10 00&-1.76&14.39$\pm$0.03&13.32$\pm$0.03&12.86$\pm$0.03&12.59$\pm$0.02&12.29$\pm$0.02&10.33$\pm$0.09&7.62$\pm$0.13&\ldots&\ldots&\ldots\\                        
J090218.98-443939.9& 09 02 19.0  &-44 39 40&-0.76&15.06$\pm$0.04&14.21$\pm$0.05&13.79$\pm$0.05&13.01$\pm$0.03&12.37$\pm$0.03&9.94$\pm$0.07&7.76$\pm$0.15&\ldots&\ldots&\ldots\\                         
J090222.99-443113.0& 09 02 23.0  &-44 31 13&-1.82&14.04$\pm$0.03&13.28$\pm$0.03&12.89$\pm$0.03&12.58$\pm$0.02&12.08$\pm$0.02&10.64$\pm$0.11&8.34(u)&\ldots&\ldots&\ldots\\                              
J090240.20-445436.1& 09 02 40.2  &-44 54 36&-1.67&15.68$\pm$0.08&13.08$\pm$0.03&11.80$\pm$0.03&11.33$\pm$0.02&11.03$\pm$0.02&9.84$\pm$0.17&7.22$\pm$0.10&\ldots&\ldots&\ldots\\                         
J090305.69-444601.7& 09 03 05.7  &-44 46 02&-1.79&13.27$\pm$0.03&12.65$\pm$0.03&12.40$\pm$0.02&12.23$\pm$0.02&12.22$\pm$0.02&9.74$\pm$0.06&7.07$\pm$0.09&\ldots&\ldots&\ldots\\                         
J090352.07-432534.2& 09 03 52.1  &-43 25 34&-0.91&\ldots&\ldots&\ldots&12.83$\pm$0.02&12.52$\pm$0.02&10.16$\pm$0.07&7.89$\pm$0.18&\ldots&\ldots&\ldots\\                                                
\enddata
\end{deluxetable}

\clearpage

\begin{deluxetable}{cccccccccc}
\rotate
\centering
\tabletypesize{\scriptsize}
\tablewidth{0pt}
\setlength{\tabcolsep}{0.2cm}
\tablecaption{New protostellar core candidates in Vela C.\label{blastproto}}
\tablehead{
\colhead{}&
\colhead{}&
\colhead{}&
\multicolumn{4}{c}{WISE}&
\multicolumn{3}{c}{BLAST}\\
\cline{4-10}
\colhead{BLAST}&
\colhead{RA}&
\colhead{DEC}&
\colhead{W1}&
\colhead{W2}&
\colhead{W3}&
\colhead{W4}&
\colhead{F250}&
\colhead{F350}&
\colhead{F500}\\
\colhead{}&
\colhead{(J2000)}&
\colhead{(J2000)}&
\colhead{(mag)}&
\colhead{(mag)}&
\colhead{(mag)}&
\colhead{(mag)}&
\colhead{(Jy)}&
\colhead{(Jy)}&
\colhead{(Jy)}
}
\startdata
J085808-424921& 08 58 08.4  &-42 49 21&11.79$\pm$0.03&10.11$\pm$0.02&8.82$\pm$0.11&6.49$\pm$0.24&62.5$\pm$9.8&43.5$\pm$8.0&20.0$\pm$5.3\\                                                               
J090016-443850& 09 00 16.2  &-44 38 50&13.04$\pm$0.03&11.82$\pm$0.02&11.13$\pm$0.18&7.46$\pm$0.15&3.2$\pm$0.5&3.3$\pm$0.5&1.9$\pm$0.3\\                                                                 
J090034-450525& 09 00 34.1  &-45 05 26&15.05$\pm$0.04&13.17$\pm$0.03&11.93(u)&8.08$\pm$0.24&22.5$\pm$3.2&18.6$\pm$2.9&9.9$\pm$1.9\\                                                                     
J090125-444804& 09 01 25.7  &-44 48 05&15.64$\pm$0.04&15.41$\pm$0.09&9.93$\pm$0.14&7.70$\pm$0.16&6.7$\pm$1.0&6.9$\pm$1.0&4.3$\pm$0.7\\                                                                  
\enddata
\end{deluxetable}

\clearpage

\begin{deluxetable}{lcclccc}
\tabletypesize{\scriptsize}
\tablecolumns{7}
\tablewidth{0pc}
\tablecaption{Possible driving sources of HH objects and EGOs in Vela C \label{extable}}
\tablehead{
\colhead{Exciting}&
\colhead{$\alpha$}&
\colhead{$\delta$}&
\colhead{Associated}&
\colhead{$l$\tablenotemark{a}}&
\colhead{PA\tablenotemark{b}}&
\colhead{}\\
\colhead{Source}&
\colhead{(J2000)}&
\colhead{(J2000)}&
\colhead{HHs and EGOs}&
\colhead{(pc)}&
\colhead{(\degr)}&
\colhead{Type}
}
\startdata
ESO-Halpha274& 08 57 49.0  &-42 37 45&HH1090,EGO2&0.35&68   &unipolar\\
ESO-Halpha2431& 08 59 01.0  &-43 00 08&HH1091&0.20&18   &unipolar\\
IRAS08569-4230& 08 58 46.5  &-42 42 24&HH1092-1094,EGO4&2.39&115  &unipolar\\
BLASTJ090016-443850& 09 00 16.2  &-44 38 50&HH1095,EGO6&0.15&136  &unipolar\\
MSXG265.9151+01.0702& 09 00 39.9  &-44 35 11&HH1096&0.30&91   &unipolar\\
MSXG265.4865+01.3474& 09 00 15.4  &-44 04 52&HH1097&1.36&146  &unipolar\\
IRS35& 09 00 38.1  &-43 59 34&HH1098&0.86&177  &unipolar\\
MSXG266.3267+00.9389& 09 01 38.9  &-44 58 50&HH1101,EGO10&0.29&168  &unipolar\\
HH74IRS& 09 02 15.8  &-44 50 12&HH1107,HH74&0.31&171  &bipolar\\
IRS76& 08 59 53.4  &-43 56 13&EGO5&0.96&38   &unipolar\\
\enddata
\tablecomments{Units of right ascension are hours, minutes, and seconds, and units of declination are degrees, arcminutes, and arcseconds.}
\tablenotetext{a}{Jet length. For unipolar outflows, it is the maximal value of length from the exciting source to the associated HH features and EGO features. For bipolar outflows, it is the accumulation of the lengths of two lobes.}
\tablenotetext{b}{Mean value of position angles of HH features and EGO features relative to their possible driving source, measured east to north. Note that we have transformed the position angles into the range of 0\degr-180\degr~disregarding the outflow is a bipolar or unipolar outflow.}
\end{deluxetable}

\clearpage
\clearpage

\clearpage

\clearpage

\clearpage

\clearpage

\clearpage

\clearpage

\clearpage

\clearpage

\clearpage

\clearpage


\clearpage

\clearpage

\clearpage

\begin{thebibliography}{}
\expandafter\ifx\csname natexlab\endcsname\relax\def\natexlab#1{#1}\fi

\bibitem[{{Aniano} {et~al.}(2011){Aniano}, {Draine}, {Gordon}, \&
  {Sandstrom}}]{kernels}
{Aniano}, G., {Draine}, B.~T., {Gordon}, K.~D., \& {Sandstrom}, K. 2011, \pasp,
  123, 1218

\bibitem[{{Arce} {et~al.}(2007){Arce}, {Shepherd}, {Gueth}, {Lee}, {Bachiller},
  {Rosen}, \& {Beuther}}]{arc07}
{Arce}, H.~G., {Shepherd}, D., {Gueth}, F., {et~al.} 2007, Protostars and
  Planets V, 245

\bibitem[{{Ayala} {et~al.}(2000){Ayala}, {Noriega-Crespo}, {Garnavich},
  {Curiel}, {Raga}, {B{\"o}hm}, \& {Raymond}}]{aya00}
{Ayala}, S., {Noriega-Crespo}, A., {Garnavich}, P.~M., {et~al.} 2000, \aj, 120,
  909

\bibitem[{{Baade} {et~al.}(1999){Baade}, {Meisenheimer}, {Iwert}, {Alonso},
  {Augusteijn}, {Beletic}, {Bellemann}, {Benesch}, {B{\"o}hm}, {B{\"o}hnhardt},
  {Brewer}, {Deiries}, {Delabre}, {Donaldson}, {Dupuy}, {Franke}, {Gerdes},
  {Gilliotte}, {Grimm}, {Haddad}, {Hess}, {Ihle}, {Klein}, {Lenzen}, {Lizon},
  {Mancini}, {M{\"u}nch}, {Pizarro}, {Prado}, {Rahmer}, {Reyes}, {Richardson},
  {Robledo}, {Sanchez}, {Silber}, {Sinclaire}, {Wackermann}, \&
  {Zaggia}}]{baade99}
{Baade}, D., {Meisenheimer}, K., {Iwert}, O., {et~al.} 1999, The Messenger, 95,
  15

\bibitem[{{Baba} {et~al.}(2006){Baba}, {Sato}, {Nagashima}, {Nishiyama},
  {Kato}, {Haba}, {Nagata}, {Nagayama}, {Tamura}, \& {Sugitani}}]{baba06}
{Baba}, D., {Sato}, S., {Nagashima}, C., {et~al.} 2006, \aj, 132, 1692

\bibitem[{{Bally} {et~al.}(2007){Bally}, {Reipurth}, \& {Davis}}]{bal07}
{Bally}, J., {Reipurth}, B., \& {Davis}, C.~J. 2007, Protostars and Planets V,
  215

\bibitem[{{Bertin} \& {Arnouts}(1996)}]{sex}
{Bertin}, E., \& {Arnouts}, S. 1996, \aaps, 117, 393

\bibitem[{{Bik} {et~al.}(2005){Bik}, {Kaper}, {Hanson}, \& {Smits}}]{bik05}
{Bik}, A., {Kaper}, L., {Hanson}, M.~M., \& {Smits}, M. 2005, \aap, 440, 121

\bibitem[{{Bik} \& {Thi}(2004)}]{bt04}
{Bik}, A., \& {Thi}, W.~F. 2004, \aap, 427, L13

\bibitem[{{Chambers} {et~al.}(2009){Chambers}, {Jackson}, {Rathborne}, \&
  {Simon}}]{cham09}
{Chambers}, E.~T., {Jackson}, J.~M., {Rathborne}, J.~M., \& {Simon}, R. 2009,
  \apjs, 181, 360

\bibitem[{{Cyganowski} {et~al.}(2008){Cyganowski}, {Whitney}, {Holden},
  {Braden}, {Brogan}, {Churchwell}, {Indebetouw}, {Watson}, {Babler},
  {Benjamin}, {Gomez}, {Meade}, {Povich}, {Robitaille}, \& {Watson}}]{ego08}
{Cyganowski}, C.~J., {Whitney}, B.~A., {Holden}, E., {et~al.} 2008, \aj, 136,
  2391

\bibitem[{{Davis} {et~al.}(2010){Davis}, {Gell}, {Khanzadyan}, {Smith}, \&
  {Jenness}}]{davis10}
{Davis}, C.~J., {Gell}, R., {Khanzadyan}, T., {Smith}, M.~D., \& {Jenness}, T.
  2010, \aap, 511, A24+

\bibitem[{{Davis} {et~al.}(2007){Davis}, {Kumar}, {Sandell}, {Froebrich},
  {Smith}, \& {Currie}}]{davis07}
{Davis}, C.~J., {Kumar}, M.~S.~N., {Sandell}, G., {et~al.} 2007, \mnras, 374,
  29

\bibitem[{{Davis} {et~al.}(2008){Davis}, {Scholz}, {Lucas}, {Smith}, \&
  {Adamson}}]{davis08}
{Davis}, C.~J., {Scholz}, P., {Lucas}, P., {Smith}, M.~D., \& {Adamson}, A.
  2008, \mnras, 387, 954

\bibitem[{{Davis} {et~al.}(2009){Davis}, {Froebrich}, {Stanke}, {Megeath},
  {Kumar}, {Adamson}, {Eisl{\"o}ffel}, {Gredel}, {Khanzadyan}, {Lucas},
  {Smith}, \& {Varricatt}}]{davis09}
{Davis}, C.~J., {Froebrich}, D., {Stanke}, T., {et~al.} 2009, \aap, 496, 153

\bibitem[{{Devlin}(2001)}]{devlin01}
{Devlin}, M. 2001, in Deep Millimeter Surveys: Implications for Galaxy
  Formation and Evolution, ed. J.~D. {Lowenthal} \& D.~H. {Hughes}, 59

\bibitem[{{Dobashi}(2011)}]{dobashi11}
{Dobashi}, K. 2011, \pasj, 63, 1

\bibitem[{{Elia} {et~al.}(2007){Elia}, {Massi}, {Strafella}, {De Luca},
  {Giannini}, {Lorenzetti}, {Nisini}, {Campeggio}, \& {Maiolo}}]{elia07}
{Elia}, D., {Massi}, F., {Strafella}, F., {et~al.} 2007, \apj, 655, 316

\bibitem[{{Ellerbroek} {et~al.}(2011){Ellerbroek}, {Kaper}, {Bik}, {de Koter},
  {Horrobin}, {Puga}, {Sana}, \& {Waters}}]{eller11}
{Ellerbroek}, L.~E., {Kaper}, L., {Bik}, A., {et~al.} 2011, \apjl, 732, L9

\bibitem[{{Ellerbroek} {et~al.}(2013){Ellerbroek}, {Podio}, {Kaper}, {Sana},
  {Huppenkothen}, {de Koter}, \& {Monaco}}]{eller13}
{Ellerbroek}, L.~E., {Podio}, L., {Kaper}, L., {et~al.} 2013, \aap, 551, A5

\bibitem[{{Fazio} {et~al.}(2004){Fazio}, {Hora}, {Allen}, {Ashby}, {Barmby},
  {Deutsch}, {Huang}, {Kleiner}, {Marengo}, {Megeath}, {Melnick}, {Pahre},
  {Patten}, {Polizotti}, {Smith}, {Taylor}, {Wang}, {Willner}, {Hoffmann},
  {Pipher}, {Forrest}, {McMurty}, {McCreight}, {McKelvey}, {McMurray}, {Koch},
  {Moseley}, {Arendt}, {Mentzell}, {Marx}, {Losch}, {Mayman}, {Eichhorn},
  {Krebs}, {Jhabvala}, {Gezari}, {Fixsen}, {Flores}, {Shakoorzadeh}, {Jungo},
  {Hakun}, {Workman}, {Karpati}, {Kichak}, {Whitley}, {Mann}, {Tollestrup},
  {Eisenhardt}, {Stern}, {Gorjian}, {Bhattacharya}, {Carey}, {Nelson},
  {Glaccum}, {Lacy}, {Lowrance}, {Laine}, {Reach}, {Stauffer}, {Surace},
  {Wilson}, {Wright}, {Hoffman}, {Domingo}, \& {Cohen}}]{irac}
{Fazio}, G.~G., {Hora}, J.~L., {Allen}, L.~E., {et~al.} 2004, \apjs, 154, 10

\bibitem[{{Flaherty} {et~al.}(2007){Flaherty}, {Pipher}, {Megeath}, {Winston},
  {Gutermuth}, {Muzerolle}, {Allen}, \& {Fazio}}]{flaherty07}
{Flaherty}, K.~M., {Pipher}, J.~L., {Megeath}, S.~T., {et~al.} 2007, \apj, 663,
  1069

\bibitem[{{G{\aa}lfalk} \& {Olofsson}(2007)}]{gal07}
{G{\aa}lfalk}, M., \& {Olofsson}, G. 2007, \aap, 466, 579

\bibitem[{{Giannini} {et~al.}(2005){Giannini}, {Massi}, {Podio}, {Lorenzetti},
  {Nisini}, {Caratti o Garatti}, {Liseau}, {Lo Curto}, \& {Vitali}}]{gian05}
{Giannini}, T., {Massi}, F., {Podio}, L., {et~al.} 2005, \aap, 433, 941

\bibitem[{{Giannini} {et~al.}(2007){Giannini}, {Lorenzetti}, {De Luca},
  {Nisini}, {Marengo}, {Allen}, {Smith}, {Fazio}, {Massi}, {Elia}, \&
  {Strafella}}]{gian07}
{Giannini}, T., {Lorenzetti}, D., {De Luca}, M., {et~al.} 2007, \apj, 671, 470

\bibitem[{{Giannini} {et~al.}(2012){Giannini}, {Elia}, {Lorenzetti},
  {Molinari}, {Motte}, {Schisano}, {Pezzuto}, {Pestalozzi}, {di Giorgio},
  {Andr{\'e}}, {Hill}, {Benedettini}, {Bontemps}, {di Francesco}, {Fallscheer},
  {Hennemann}, {Kirk}, {Minier}, {Nguyen Luong}, {Polychroni}, {Rygl},
  {Saraceno}, {Schneider}, {Spinoglio}, {Testi}, {Ward-Thompson}, \&
  {White}}]{gian12}
{Giannini}, T., {Elia}, D., {Lorenzetti}, D., {et~al.} 2012, \aap, 539, A156

\bibitem[{{Giannini} {et~al.}(2013){Giannini}, {Lorenzetti}, {De Luca},
  {Strafella}, {Elia}, {Maiolo}, {Marengo}, {Maruccia}, {Massi}, {Nisini},
  {Olmi}, {Salama}, \& {Smith}}]{gian13}
{Giannini}, T., {Lorenzetti}, D., {De Luca}, M., {et~al.} 2013, \apj, 767, 147

\bibitem[{{Greene} {et~al.}(1994){Greene}, {Wilking}, {Andre}, {Young}, \&
  {Lada}}]{greene94}
{Greene}, T.~P., {Wilking}, B.~A., {Andre}, P., {Young}, E.~T., \& {Lada},
  C.~J. 1994, \apj, 434, 614

\bibitem[{{Gutermuth} {et~al.}(2009){Gutermuth}, {Megeath}, {Myers}, {Allen},
  {Pipher}, \& {Fazio}}]{gutermuth09}
{Gutermuth}, R.~A., {Megeath}, S.~T., {Myers}, P.~C., {et~al.} 2009, \apjs,
  184, 18

\bibitem[{{Gutermuth} {et~al.}(2008){Gutermuth}, {Myers}, {Megeath}, {Allen},
  {Pipher}, {Muzerolle}, {Porras}, {Winston}, \& {Fazio}}]{gutermuth08}
{Gutermuth}, R.~A., {Myers}, P.~C., {Megeath}, S.~T., {et~al.} 2008, \apj, 674,
  336

\bibitem[{{Hartigan} {et~al.}(1999){Hartigan}, {Morse}, {Tumlinson}, {Raymond},
  \& {Heathcote}}]{hartigan99}
{Hartigan}, P., {Morse}, J.~A., {Tumlinson}, J., {Raymond}, J., \& {Heathcote},
  S. 1999, \apj, 512, 901

\bibitem[{{Harvey} {et~al.}(2006){Harvey}, {Chapman}, {Lai}, {Evans}, {Allen},
  {J{\o}rgensen}, {Mundy}, {Huard}, {Porras}, {Cieza}, {Myers}, {Mer{\'{\i}}n},
  {van Dishoeck}, {Young}, {Spiesman}, {Blake}, {Koerner}, {Padgett},
  {Sargent}, \& {Stapelfeldt}}]{harvey06}
{Harvey}, P.~M., {Chapman}, N., {Lai}, S.-P., {et~al.} 2006, \apj, 644, 307

\bibitem[{{Hill} {et~al.}(2011){Hill}, {Motte}, {Didelon}, {Bontemps},
  {Minier}, {Hennemann}, {Schneider}, {Andr{\'e}}, {Men'shchikov}, {Anderson},
  {Arzoumanian}, {Bernard}, {di Francesco}, {Elia}, {Giannini}, {Griffin},
  {K{\"o}nyves}, {Kirk}, {Marston}, {Martin}, {Molinari}, {Nguyen Luong},
  {Peretto}, {Pezzuto}, {Roussel}, {Sauvage}, {Sousbie}, {Testi},
  {Ward-Thompson}, {White}, {Wilson}, \& {Zavagno}}]{hill11}
{Hill}, T., {Motte}, F., {Didelon}, P., {et~al.} 2011, \aap, 533, A94

\bibitem[{{Hollenbach} \& {McKee}(1989)}]{hollenbach89}
{Hollenbach}, D., \& {McKee}, C.~F. 1989, \apj, 342, 306

\bibitem[{{Jarrett} {et~al.}(2011){Jarrett}, {Cohen}, {Masci}, {Wright},
  {Stern}, {Benford}, {Blain}, {Carey}, {Cutri}, {Eisenhardt}, {Lonsdale},
  {Mainzer}, {Marsh}, {Padgett}, {Petty}, {Ressler}, {Skrutskie}, {Stanford},
  {Surace}, {Tsai}, {Wheelock}, \& {Yan}}]{jarrett11}
{Jarrett}, T.~H., {Cohen}, M., {Masci}, F., {et~al.} 2011, \apj, 735, 112

\bibitem[{{Koenig} {et~al.}(2012){Koenig}, {Leisawitz}, {Benford}, {Rebull},
  {Padgett}, \& {Assef}}]{koenig12}
{Koenig}, X.~P., {Leisawitz}, D.~T., {Benford}, D.~J., {et~al.} 2012, \apj,
  744, 130

\bibitem[{{Konigl} \& {Pudritz}(2000)}]{kon00}
{Konigl}, A., \& {Pudritz}, R.~E. 2000, Protostars and Planets IV, 759

\bibitem[{{Kwitter} \& {Henry}(1998)}]{kwitter98}
{Kwitter}, K.~B., \& {Henry}, R.~B.~C. 1998, \apj, 493, 247

\bibitem[{{Kwitter} \& {Henry}(2001)}]{kwitter01}
---. 2001, \apj, 562, 804

\bibitem[{{Liseau} {et~al.}(1992){Liseau}, {Lorenzetti}, {Nisini}, {Spinoglio},
  \& {Moneti}}]{lis92}
{Liseau}, R., {Lorenzetti}, D., {Nisini}, B., {Spinoglio}, L., \& {Moneti}, A.
  1992, \aap, 265, 577

\bibitem[{{Lorenzetti} {et~al.}(2002){Lorenzetti}, {Giannini}, {Vitali},
  {Massi}, \& {Nisini}}]{loren02}
{Lorenzetti}, D., {Giannini}, T., {Vitali}, F., {Massi}, F., \& {Nisini}, B.
  2002, \apj, 564, 839

\bibitem[{{Massi} {et~al.}(2013){Massi}, {Giannini}, {Lorenzetti}, {Strafella},
  {Elia}, {Olmi}, {Maruccia}, \& {De Luca}}]{massiposter}
{Massi}, F., {Giannini}, T., {Lorenzetti}, D., {et~al.} 2013, in Protostars and
  Planets VI, Heidelberg, July 15-20, 2013. Poster \#1B010, 10

\bibitem[{{Massi} {et~al.}(2003){Massi}, {Lorenzetti}, \& {Giannini}}]{massi03}
{Massi}, F., {Lorenzetti}, D., \& {Giannini}, T. 2003, \aap, 399, 147

\bibitem[{{Massi} {et~al.}(2000){Massi}, {Lorenzetti}, {Giannini}, \&
  {Vitali}}]{massi00}
{Massi}, F., {Lorenzetti}, D., {Giannini}, T., \& {Vitali}, F. 2000, \aap, 353,
  598

\bibitem[{{Molinari} {et~al.}(1993){Molinari}, {Liseau}, \&
  {Lorenzetti}}]{molinari93}
{Molinari}, S., {Liseau}, R., \& {Lorenzetti}, D. 1993, \aaps, 101, 59

\bibitem[{{Murphy} \& {May}(1991)}]{mm91}
{Murphy}, D.~C., \& {May}, J. 1991, \aap, 247, 202

\bibitem[{{Netterfield} {et~al.}(2009){Netterfield}, {Ade}, {Bock}, {Chapin},
  {Devlin}, {Griffin}, {Gundersen}, {Halpern}, {Hargrave}, {Hughes}, {Klein},
  {Marsden}, {Martin}, {Mauskopf}, {Olmi}, {Pascale}, {Patanchon}, {Rex},
  {Roy}, {Scott}, {Semisch}, {Thomas}, {Truch}, {Tucker}, {Tucker}, {Viero}, \&
  {Wiebe}}]{netterfield09}
{Netterfield}, C.~B., {Ade}, P.~A.~R., {Bock}, J.~J., {et~al.} 2009, \apj, 707,
  1824

\bibitem[{{Noriega-Crespo} {et~al.}(2004){Noriega-Crespo}, {Morris}, {Marleau},
  {Carey}, {Boogert}, {van Dishoeck}, {Evans}, {Keene}, {Muzerolle},
  {Stapelfeldt}, {Pontoppidan}, {Lowrance}, {Allen}, \& {Bourke}}]{nc04}
{Noriega-Crespo}, A., {Morris}, P., {Marleau}, F.~R., {et~al.} 2004, \apjs,
  154, 352

\bibitem[{{Pascale} {et~al.}(2008){Pascale}, {Ade}, {Bock}, {Chapin}, {Chung},
  {Devlin}, {Dicker}, {Griffin}, {Gundersen}, {Halpern}, {Hargrave}, {Hughes},
  {Klein}, {MacTavish}, {Marsden}, {Martin}, {Martin}, {Mauskopf},
  {Netterfield}, {Olmi}, {Patanchon}, {Rex}, {Scott}, {Semisch}, {Thomas},
  {Truch}, {Tucker}, {Tucker}, {Viero}, \& {Wiebe}}]{pascale08}
{Pascale}, E., {Ade}, P.~A.~R., {Bock}, J.~J., {et~al.} 2008, \apj, 681, 400

\bibitem[{{Petterson}(2008)}]{pettersson08}
{Petterson}, B. 2008, in Handbook of Star Forming Regions. vol. 2, 43-123
  (2008), 43--123

\bibitem[{{Pettersson} \& {Reipurth}(1994)}]{pre94}
{Pettersson}, B., \& {Reipurth}, B. 1994, \aaps, 104, 233

\bibitem[{{Qiu} {et~al.}(2008){Qiu}, {Zhang}, {Megeath}, {Gutermuth},
  {Beuther}, {Shepherd}, {Sridharan}, {Testi}, \& {De Pree}}]{qiu08}
{Qiu}, K., {Zhang}, Q., {Megeath}, S.~T., {et~al.} 2008, \apj, 685, 1005

\bibitem[{{Reach} {et~al.}(2006){Reach}, {Rho}, {Tappe}, {Pannuti}, {Brogan},
  {Churchwell}, {Meade}, {Babler}, {Indebetouw}, \& {Whitney}}]{reach06}
{Reach}, W.~T., {Rho}, J., {Tappe}, A., {et~al.} 2006, \aj, 131, 1479

\bibitem[{{Reipurth}(1981)}]{re6}
{Reipurth}, B. 1981, \aaps, 44, 379

\bibitem[{{Reipurth} \& {Bally}(2001)}]{rei01}
{Reipurth}, B., \& {Bally}, J. 2001, \araa, 39, 403

\bibitem[{{Reipurth} \& {Graham}(1988)}]{reipurth88}
{Reipurth}, B., \& {Graham}, J.~A. 1988, \aap, 202, 219

\bibitem[{{Riera} {et~al.}(2001){Riera}, {L{\'o}pez}, {Raga}, {Anglada}, \&
  {Estalella}}]{riera01}
{Riera}, A., {L{\'o}pez}, R., {Raga}, A.~C., {Anglada}, G., \& {Estalella}, R.
  2001, \rmxaa, 37, 147

\bibitem[{{Rosenthal} {et~al.}(2000){Rosenthal}, {Bertoldi}, \&
  {Drapatz}}]{rosen00}
{Rosenthal}, D., {Bertoldi}, F., \& {Drapatz}, S. 2000, \aap, 356, 705

\bibitem[{{Shang} {et~al.}(2007){Shang}, {Li}, \& {Hirano}}]{sha07}
{Shang}, H., {Li}, Z.-Y., \& {Hirano}, N. 2007, Protostars and Planets V, 261

\bibitem[{{Shu} {et~al.}(1994){Shu}, {Najita}, {Ostriker}, {Wilkin}, {Ruden},
  \& {Lizano}}]{shu94}
{Shu}, F., {Najita}, J., {Ostriker}, E., {et~al.} 1994, \apj, 429, 781

\bibitem[{{Shu} {et~al.}(2000){Shu}, {Najita}, {Shang}, \& {Li}}]{shu00}
{Shu}, F.~H., {Najita}, J.~R., {Shang}, H., \& {Li}, Z.-Y. 2000, Protostars and
  Planets IV, 789

\bibitem[{{Skrutskie} {et~al.}(2006){Skrutskie}, {Cutri}, {Stiening},
  {Weinberg}, {Schneider}, {Carpenter}, {Beichman}, {Capps}, {Chester},
  {Elias}, {Huchra}, {Liebert}, {Lonsdale}, {Monet}, {Price}, {Seitzer},
  {Jarrett}, {Kirkpatrick}, {Gizis}, {Howard}, {Evans}, {Fowler}, {Fullmer},
  {Hurt}, {Light}, {Kopan}, {Marsh}, {McCallon}, {Tam}, {Van Dyk}, \&
  {Wheelock}}]{2mass}
{Skrutskie}, M.~F., {Cutri}, R.~M., {Stiening}, R., {et~al.} 2006, \aj, 131,
  1163

\bibitem[{{Strafella} {et~al.}(2010){Strafella}, {Elia}, {Campeggio},
  {Giannini}, {Lorenzetti}, {Marengo}, {Smith}, {Fazio}, {De Luca}, \&
  {Massi}}]{stra10}
{Strafella}, F., {Elia}, D., {Campeggio}, L., {et~al.} 2010, \apj, 719, 9

\bibitem[{{Teixeira} {et~al.}(2008){Teixeira}, {McCoey}, {Fich}, \&
  {Lada}}]{tei08}
{Teixeira}, P.~S., {McCoey}, C., {Fich}, M., \& {Lada}, C.~J. 2008, \mnras,
  384, 71

\bibitem[{{Thi} \& {Bik}(2005)}]{tb05}
{Thi}, W.-F., \& {Bik}, A. 2005, \aap, 438, 557

\bibitem[{{Wang} \& {Henning}(2006)}]{wh06}
{Wang}, H., \& {Henning}, T. 2006, \apj, 643, 985

\bibitem[{{Wang} \& {Henning}(2009)}]{wh09}
---. 2009, \aj, 138, 1072

\bibitem[{{Wang} {et~al.}(2004){Wang}, {Mundt}, {Henning}, \& {Apai}}]{wang04}
{Wang}, H., {Mundt}, R., {Henning}, T., \& {Apai}, D. 2004, \apj, 617, 1191

\bibitem[{{Wang} {et~al.}(2001){Wang}, {Yang}, {Wang}, {Deng}, {Yan}, \&
  {Chen}}]{wang01}
{Wang}, H., {Yang}, J., {Wang}, M., {et~al.} 2001, \aj, 121, 1551

\bibitem[{{Wang} {et~al.}(2003){Wang}, {Yang}, {Wang}, \& {Yan}}]{wang03}
{Wang}, H., {Yang}, J., {Wang}, M., \& {Yan}, J. 2003, \aj, 125, 842

\bibitem[{{Wouterloot} \& {Brand}(1999)}]{wb99}
{Wouterloot}, J.~G.~A., \& {Brand}, J. 1999, \aaps, 140, 177

\bibitem[{{Wright} {et~al.}(2010){Wright}, {Eisenhardt}, {Mainzer}, {Ressler},
  {Cutri}, {Jarrett}, {Kirkpatrick}, {Padgett}, {McMillan}, {Skrutskie},
  {Stanford}, {Cohen}, {Walker}, {Mather}, {Leisawitz}, {Gautier}, {McLean},
  {Benford}, {Lonsdale}, {Blain}, {Mendez}, {Irace}, {Duval}, {Liu}, {Royer},
  {Heinrichsen}, {Howard}, {Shannon}, {Kendall}, {Walsh}, {Larsen}, {Cardon},
  {Schick}, {Schwalm}, {Abid}, {Fabinsky}, {Naes}, \& {Tsai}}]{wise}
{Wright}, E.~L., {Eisenhardt}, P.~R.~M., {Mainzer}, A.~K., {et~al.} 2010, \aj,
  140, 1868

\bibitem[{{Yamaguchi} {et~al.}(1999){Yamaguchi}, {Mizuno}, {Saito},
  {Matsunaga}, {Mizuno}, {Ogawa}, \& {Fukui}}]{yam99b}
{Yamaguchi}, N., {Mizuno}, N., {Saito}, H., {et~al.} 1999, \pasj, 51, 775

\bibitem[{{Zacharias} {et~al.}(2013){Zacharias}, {Finch}, {Girard}, {Henden},
  {Bartlett}, {Monet}, \& {Zacharias}}]{ucac4}
{Zacharias}, N., {Finch}, C.~T., {Girard}, T.~M., {et~al.} 2013, \aj, 145, 44

\bibitem[{{Zhang} \& {Wang}(2009)}]{zw09}
{Zhang}, M., \& {Wang}, H. 2009, \aj, 138, 1830

\bibitem[{{Zhang} {et~al.}(2013){Zhang}, {Brandner}, {Wang}, {Gennaro}, {Bik},
  {Henning}, {Gredel}, {Smith}, \& {Stanke}}]{zhang13}
{Zhang}, M., {Brandner}, W., {Wang}, H., {et~al.} 2013, \aap, 553, A41

\end{thebibliography}
\end{document}